\documentclass[english,12pt,a4paper,openany]{article}

\usepackage{graphicx}
\usepackage{dcolumn}
\usepackage{bm}
\usepackage{braket}
\usepackage{authblk}
\usepackage{hyperref}
\usepackage{amsthm}
\usepackage{amssymb}
\usepackage{xcolor}
\usepackage{amsmath}

\makeatletter
\def\@fnsymbol#1{\ensuremath{\ifcase#1\or \dagger\or \ddagger\or *\or
		\mathsection\or \mathparagraph\or |\or **\or \dagger\dagger
		\or \ddagger\ddagger \else\@ctrerr\fi}}
\makeatother
\usepackage[backend=bibtex,style=nature]{biblatex}
\bibliography{non_herm}
\usepackage{cleveref}
\begin{document}
\date{}
\title{Non-Hermitian quantum gases: a platform for imaginary time crystals}


\author[1,2,3]{R. Arouca\thanks{rodrigo.arouca@physics.uu.se, corresponding author.}}
\author[2]{E. C. Marino\thanks{marino@if.ufrj.br.}}
\author[1]{C. Morais Smith\thanks{C.deMoraisSmith@uu.nl.}}
\affil[1]{\small{Institute for Theoretical Physics, Utrecht University, Princetonplein 5, 3584 CC Utrecht, The Netherlands}}
\affil[2]{\small{Instituto de F\' isica, Universidade Federal do Rio de Janeiro, C.P. 68528, Rio de Janeiro, RJ, 21941-972, Brazil}}
\affil[3]{\small{Department of Physics and Astronomy, Uppsala University, Box 516, S-751 20 Uppsala, Sweden}}%

\maketitle

\begin{abstract}
	One of the foremost objectives of statistical mechanics is the description of the thermodynamic properties of quantum gases. Despite the great importance of this topic, such achievement is still lacking in the case of non-Hermitian quantum gases. Here, we investigate the properties of bosonic and fermionic non-Hermitian systems at finite temperatures. We show that non-Hermitian systems exhibit oscillations both in temperature and imaginary time. As such, they can be a possible platform to realize an imaginary time crystal (iTC) phase. The Hatano-Nelson model is identified as a simple lattice model to reveal this effect. In addition, we show that the conditions for the iTC to be manifest are the same as the conditions for the presence of disorder points, where the correlation functions show oscillating behavior. This analysis makes clear that our realization of an iTC is effectively a way to filter one specific Matsubara mode. In this realization, the Matsubara frequency, which usually appears as a mathematical tool to compute correlation functions at finite temperatures, can be measured experimentally.
	
	
\end{abstract}

\section{Introduction}\label{sec_intro} 

	The evolution of a system in imaginary time $\tau$ is a long known prescription in quantum field theory to compute the partition function \cite{marino2017quantum}. In this formalism, periodic boundary conditions lead to discrete sets of frequencies, the Matsubara frequencies $\omega_n$. Despite their pivotal importance, these frequencies are believed to be just tools to compute correlation functions. Observing the structure of the Green's function $G$, one concludes that there will be no poles associated with $\omega_n$, but only to the real frequencies, with a small imaginary part related to the causal structure of $G$. A natural question emerges then when one considers systems with sizable values of the imaginary part of theirs energies, as it occurs for non-Hermitian systems.

Non-Hermitian Hamiltonians arise as an effective description of open systems \cite{moiseyev2011non, bergholtz2019exceptional, ashida2020non} and lead to novel properties that cannot be observed in a closed system \cite{ashida2020non, bergholtz2019exceptional}. Some examples are: (i) the extension of the symmetry-protected topological phases described by the Altland-Zirnbauer classification from a 10-fold to a 38-fold table due to the splitting of the usual discrete symmetries \cite{bergholtz2019exceptional, ashida2020non, kawabata2019symmetry, kawabata2019classification}; ii) the non-Hermitian skin effect (NHSE) \cite{yao2018edge,lee2019anatomy, lee2019unraveling,borgnia2020non, PhysRevLett.124.086801, zhang2020correspondence, li2020critica, helbig2020generalized, xiao2020non}, which is an accumulation of modes in one of the edges of the system; (iii) the extensive dependence of boundary conditions, such that systems presenting open boundary conditions (OBC) and periodic boundary conditions (PBC) have completely different spectrum \cite{lee2019anatomy, lee2019unraveling}; and (iv) anomalous behavior in quantum phase transitions \cite{arouca2020unconventional, matsumoto2020continuous}.

The simplest kind of non-Hermitian system is a non-Hermitian quantum gas. A quantum gas is a non-interacting system, such that its Hamiltonian $\hat{H}$ can be decomposed as the sum of Hamiltonians of some quantum numbers $m$, $\hat{H}=\oplus_{m} \hat{h}_{m}.$ The thermodynamic properties of a non-Hermitian ideal quantum gas were not explored so far; yet, in the presence of a pseudo-Hermitian symmetry, a biorthogonal thermodynamic description is available in the literature \cite{gardas2016non}. Although the thermodynamic limit is not really achievable for large OBC systems due to the NHSE, a finite-size thermodynamic analysis is still consistent and one can obtain results in the thermodynamic limit for PBC and for the surrogate Hamiltonian (SH). The latter consists of an analytical continuation of the Bloch Hamiltonian that reproduces the bulk spectrum of the system with OBC \cite{yao2018edge, gong2018topological, lee2019unraveling}. 

Here, we investigate the thermodynamic behavior of non-Hermitian quantum gases at \textit{finite temperature}. We find that for a range of intermediary temperatures, this system exhibits oscillations in both $\beta=1/(k_BT)$ and imaginary time $\tau$. This is precisely a footprint of the imaginary time crystal (iTC) phase conjectured by Wilczek in his original paper on time crystals \cite{wilczek2012quantum}. The conditions for the existence of this phase are the same as the ones for the presence of disorder points \cite{stephenson1970ising1, stephenson1970ising,stephenson1970ising, stephenson1970range, stephenson1970two, chakrabarty2011modulation, chakrabarty2012universality, timonin2021disorder}- critical phases at which the correlation function has a modulation, together with the exponential decay. We apply our results to the Hatano-Nelson model \cite{hatano1996localization, bergholtz2019exceptional} and show how one can observe these oscillating phases in both space and imaginary time.

\section{Results}

\subsection{Thermodynamics of the Hatano-Nelson model} 

We consider the thermodynamics of a non-Hermitian quantum gas with modes $m$ and energies $\epsilon_m$. One can use a biorthogonal basis to compute the partition function of these systems, see Methods, 
\begin{eqnarray}
	\mathcal{Z}_\text{B/F}&=&\prod\limits_{m}\left[1\mp\exp\left(-\beta \zeta_m\right)\right]^{\mp 1},
	\label{eq_Z_B}
\end{eqnarray}
where we introduce the subscript $B$ for bosons and $F$ for fermions, and $\zeta_m=\epsilon_m-\mu$. From $\mathcal{Z}$, one can obtain all the thermodynamic quantities, in analogy to what is done in a Hermitian system. 

Notice that these expressions have the same functional form as the ones for Hermitian gases \cite{salinas2001introduction}. However, there are important differences. The first is that the spectrum of non-Hermitian systems depends on the boundary conditions. This implies that the thermodynamic potentials will also change for different boundary conditions. In particular, the system with OBC is unstable for large system sizes, such that it can only be analyzed for small sizes. Nevertheless, the thermodynamic limit of this system can be achieved using the surrogate Hamiltonian \cite{arouca2020unconventional}, that has the same bulk spectrum. Furthermore, responses of the system, encoded in generalized forces (derivatives of the grand potential $\Omega$ with respect to the perturbation) are proportional to a biorthogonal correlation function \cite{gardas2016non}. More importantly, although the partition function is real in the presence of pseudo-Hermitian or parity-time (PT) symmetry \cite{gardas2016non} (see discussion in Methods), these complex energies change quantitatively the behavior of the thermodynamic potentials. In special, systems with $\text{Im}\zeta_m/\text{Re}\zeta_m>1$ will show oscillations in $\beta$, together with a monotonic behavior. 

\begin{figure*}[!th]
	\centering
	\includegraphics[width=\linewidth]{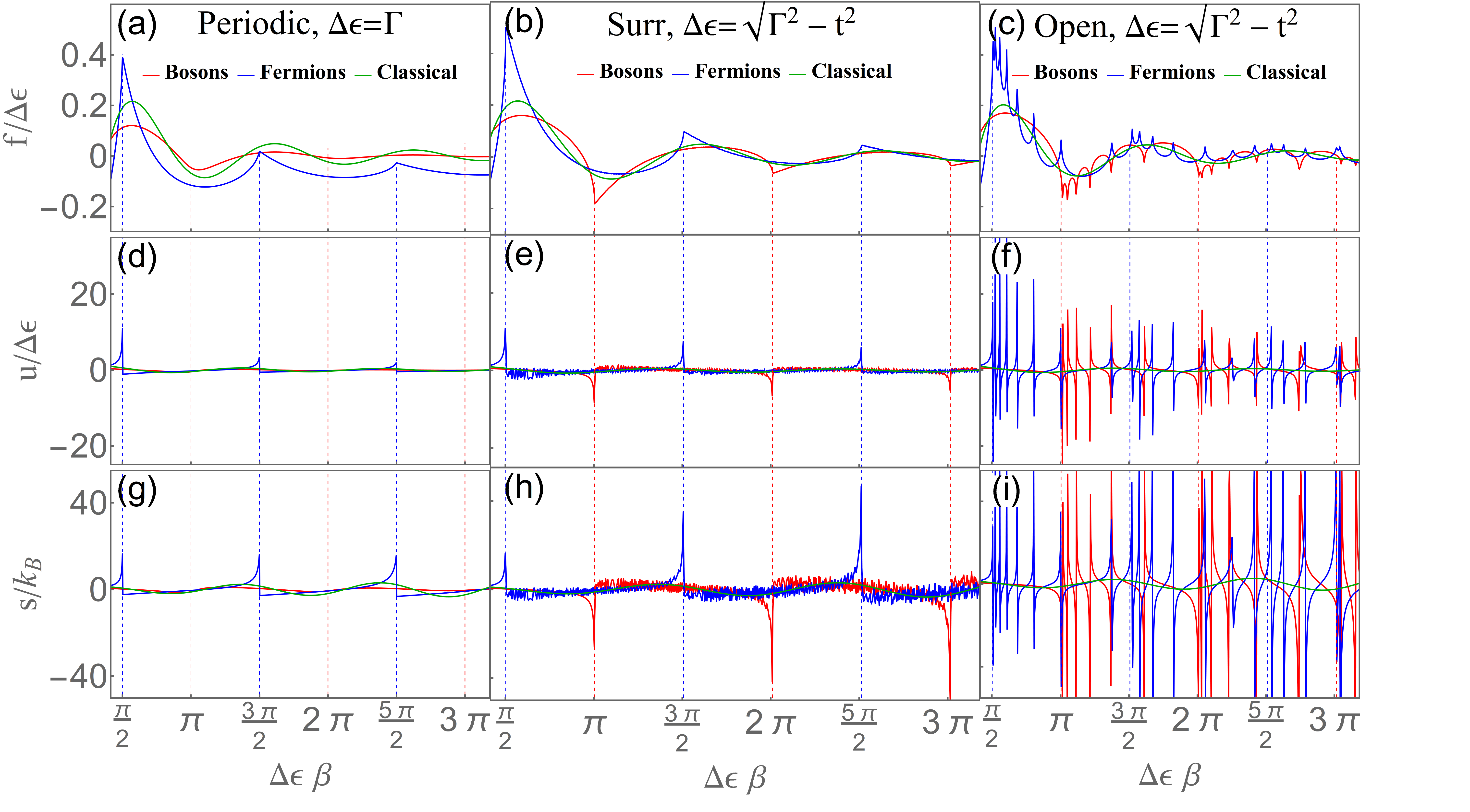}
	\caption{Thermodynamic potentials as a function of $\beta$ for the Hatano-Nelson model with $t=0.1$ for bosonic (red), fermionic (blue) and classical (green) systems. We set $\mu=-10^{-3}$, except for the bosonic system with PBC, where we use $\mu=-2t-10^{-3}$, such that the chemical potential is below the real part of the bands. For the periodic system we use $\Gamma=1$, whereas for the open and surrogate systems we use $\Gamma=\sqrt{1+t^2}$, such that the typical energy scale $\Delta \epsilon$ associated with the Hatano-Nelson model is equal to $1$ in all cases. The intensive grand potential $f$ is shown in (a) for PBC, (b) surrogate, and (c) OBC. The intensive internal energy $u$ is shown in (d) for PBC, (e) surrogate, and (f) OBC. The intensive entropy $s$ is shown in (g) for PBC, (h) surrogate, and (i) OBC. The pronunced peaks that appear occur for $2\Delta \epsilon \beta= (2n+1)\pi$ for fermions (shown with dashed blue lines in all figures) and $2\Delta \epsilon \beta= (2n)\pi$ for bosons (shown with dashed red lines in all figures). We use $1000$ $k$-points for the periodic systems (equivalent to a lattice with $1000$ sites) and a lattice with $20$ sites for the open boundary system.}
	\label{fig_HN_thermo}
\end{figure*}

To investigate this matter, we consider one paradigmatic non-Hermitian system, the Hatano-Nelson model \cite{hatano1996localization, gong2018topological}
\begin{equation}
	H=-\sum\limits_{i}^M \left[\left(t-\Gamma\right) c_j^\dagger c_{j+1}+\left(t+\Gamma\right) c_{j+1}^\dagger c_{j}\right],
\end{equation}
where $M$ is the lattice size and $c_j$ ($c_j^\dagger$) annihilates (creates) a particle at site $j$. This is a simple hopping model with a reciprocal part (proportional to $t$) and a nonreciprocal part (proportional to $\Gamma$). The behavior of the spectrum depends on $\Gamma/t$ and the boundary conditions, as discussed in detail in the Supplementary Material. In particular, for $\left|\Gamma/t\right|>1$ and OBC, the system presents a completely imaginary spectrum, thus realizing an ideal platform to observe the oscillations in $\beta$.
 The results for several thermodynamic potentials are shown in Fig.~\ref{fig_HN_thermo}. 

We consider the grand potential $\Omega$, internal energy $U$ and entropy $S$ for the bosonic, fermionic, and classical realizations of the Hatano-Nelson model, see definitions in Methods. Some care should be taken with the values chosen for the chemical potential, as for bosons, $\mu<\text{Re }\epsilon_m$ for all modes \cite{salinas2001introduction}. Although the PT-broken phase for OBC displays purely imaginary energies, and hence this is not a problem, for bosons with PBC we need to choose $\mu<-2\left|t\right|$. Consequently, only energies at the minimum of the band will show oscillatory behavior, which will decay for increasing $\beta$. Nevertheless, for the other boundary conditions and for fermionic systems, the oscillations in $\beta$ should be present, and $\mu$ will simply regularize the divergences of the thermodynamic potentials. The results are shown in Fig.~\ref{fig_HN_thermo} for the system with PBC, for the surrogate Hamiltonian, and for the system with OBC. Notice that we are using the intensive quantities that are defined by division of the extensive thermodynamic quantities by the number of sites in the lattice $M$ ($f=\Omega/M$, $u=U/M$ and $s=S/M$), such that we can analyze systems with different sizes together. For all the boundary conditions, except for bosons with PBC, the grand potential [Figs.~\ref{fig_HN_thermo} (a)-(c)], the internal energy [Figs.~\ref{fig_HN_thermo} (d)-(f)], and the entropy [Figs.~\ref{fig_HN_thermo} (g)-(i)] show oscillations as a function of $\beta$. There are slow fluctuations with the period determined by the typical energy scale $\Delta \epsilon$ of the imaginary part of the spectrum. For PBC, $\Delta\epsilon=\Gamma$, whereas for both OBC and the surrogate spectrum, $\Delta\epsilon=\sqrt{\Gamma^2-t^2}$. Hence, the period for all components and boundary conditions is given by $\pi/\Delta \epsilon$. Remarkably, the quantum systems show clear peaks for some special values of $\beta$
\begin{equation}
	\beta=\frac{\pi}{2\Delta\epsilon} \times\begin{cases} 2n, \text{bosons}\\2n+1, \text{fermions},\end{cases}
	\label{eq_beta_res}
\end{equation}
where $n \in \mathbb{Z}$.

There is, however, a noteworthy difference for the thermodynamic potentials for fermions with PBC and for bosons and fermions with OBC and the surrogate Hamiltonian. For PBC  [Figs.~\ref{fig_HN_thermo} (a), (d) and (g)], only the large scale peaks are present. This is clearly seen when inspecting the internal energy, Fig.~\ref{fig_HN_thermo} (d), or the entropy, Fig.~\ref{fig_HN_thermo} (g), which show accentuated peaks only for the values of $\beta$ described by Eq.~\eqref{eq_beta_res}. For the other boundary conditions [Figs.~\ref{fig_HN_thermo} (b), (c), (e), (f), (h) and (i)], there are fast oscillations, which although not visible in $f_{\text{surr}}$ [Figs.~\ref{fig_HN_thermo} (b)], are clearly visible in both $u_{\text{surr}}$ and $s_{\text{surr}}$ [Figs.~\ref{fig_HN_thermo} (e) and (h)], as they are both related to derivatives of $f$. The difference between both behaviors could have been anticipated because the spectrum for PBC has a non-zero real part for every $\Gamma$  (see Supplementary Material), whereas the OBC system has a purely imaginary spectrum (see Supplementary Material) for $\left|\Gamma\right|>\left|t\right|$. Then, one can expect that more oscillations will be present. 

These oscillations in $\beta$ are very interesting because $u$ is usually inversely proportional to $\beta$, such that an increasing temperature (decreasing $\beta$) would lead to a larger internal energy. Although there is a general decay of $u$ for large $\beta$, which goes to zero in the limit of $\beta\rightarrow \infty$ ($T\rightarrow 0$), a small variation of $\beta$ can lead to a large variation in the internal energy, a situation typical of the one occurring in the vicinity of a critical point \cite{salinas2001introduction}.

The values of the peaks given by Eqs.~\eqref{eq_beta_res} are the values that lead to the zeros (poles) of the fermionic (bosonic) partition function in Eq.~\eqref{eq_Z_B}, such that they describe a phase transition in the theory of Yang-Lee/Fisher zeros \cite{salinas2001introduction, yang1952statistical, lee1952statistical, bena2005statistical, tong2006lee, mussardo2017yang, timonin2021disorder, pires2021probing}. Later, we will elaborate further on this special kind of phase transition. 

The observation of oscillations in the thermodynamic quantities is not completely new. Intriguingly, features like those were also observed, although not discussed in these terms, in an analysis of a Wigner-Weyl representation of a non-reciprocal, therefore non-Hermitian, classical system \cite{nicacio2021weyl}. Similarly, the Loschmidt overlap, which is the analogous of the free energy for evolution in real time, shows similar features for Hermitian dynamical phase transitions \cite{sadrzadeh2021dynamical}. More interestingly, these oscillations are also a signature of the iTC phase proposed by Wilczek in his seminal paper on time crystals \cite{wilczek2012quantum}. Those were studied in terms of dissipative systems in Ref.~\cite{cai2020imaginary}. As we will show now, the conditions set by Eqs.~\eqref{eq_beta_res} are precisely the ones that define the iTC phase. 

\subsection{Connection to imaginary time crystals}\label{sec_iTC}

To investigate this matter, we express now the partition function as a path integral over coherent states in imaginary (or Euclidean) time $\tau$ \cite{marino2017quantum, stoof2009ultracold, altland2010condensed}. This is done using the Trotter decomposition, where we introduce an (over) complete set of coherent states at every interval in imaginary time \cite{stoof2009ultracold}. One can build such states using coherent states of a Hermitian operator, such as position, even for a non-Hermitian Hamiltonian. For completeness, we show in the Supplementary Material how to do second quantization and build coherent states for a pseudo-Hermitian operator. These states are parametrized by the fields $\Psi$ and $\Psi^\dagger$ and one can write the partition function as a path integral over them,
\begin{equation}
	\mathcal{Z}=\int D\Psi^\dagger D\Psi \, e^{-S_E[\Psi^\dagger,\Psi]},
\end{equation}
where the Euclidean action $S_E$ is given (for a quantum gas) by 
\begin{equation}
	S_E\left[\Psi^\dagger, \Psi\right]=\int\limits_{0}^{\hbar \beta}d\tau \, \Psi^\dagger\left(\tau\right)\left[\hbar\partial_\tau-\mu+\mathbb{H}\right]\Psi\left( \tau\right),
\end{equation}
with $\mathbb{H}$ the Hamiltonian matrix, such that the Hamiltonian density is $\Psi^\dagger \mathbb{H} \Psi$. These fields can depend on some continuum index, such as position or momentum, and can have also a tensorial structure accounting for spin or other inner degrees of freedom. Integration/summation/contraction over such degrees of freedom is implied. In addition, the components of $\Psi$ will be Grassmann variables if it describes a fermionic field. Because for quantum gases the action is quadratic in the fields, one can exactly integrate this theory and the thermodynamic behavior is fully determined by $\hbar\partial_\tau-\mu+\mathbb{H}$ or, equivalently, the inverse of the Green's function $G$. As a matter of fact, \cite{marino2017quantum}
\begin{equation}
	\mathcal{Z}_{B/F}=\det\left(\hbar\partial_\tau-\mu+\mathbb{H}\right)^{\mp 1}=\det\left(G\right)^{\pm 1}.
\end{equation}
Besides determining the thermodynamic behavior of the system, $G$ describes the response of the system to external perturbations. Wick theorem states \cite{marino2017quantum}  that any correlation function will be proportional to products of two-point functions, which are given by $G$. Therefore, $G$ dictates the whole behavior of a quantum gas. 

In particular, it is useful to consider the Fourier transform (FT) of $G$ in $\tau$. Due to the periodic conditions in $\tau$, the frequencies will be discrete, being even (odd) multiples of $\pi/\beta$ for bosons (fermions) \cite{stoof2009ultracold, marino2017quantum}. Therefore, we have the Matsubara frequencies $\omega_n$ 
\begin{equation}
	\hbar\omega_n=\frac{\pi}{\beta}n_M=\frac{\pi}{\beta}\begin{cases}2n, \textrm{bosons},\\\left(2n+1\right), \textrm{fermions},\end{cases}
	\label{eq_matsu}
\end{equation}
with $n_M$ the Matsubara mode. 
\begin{figure}[!h]
	\centering
	\includegraphics[width=\linewidth]{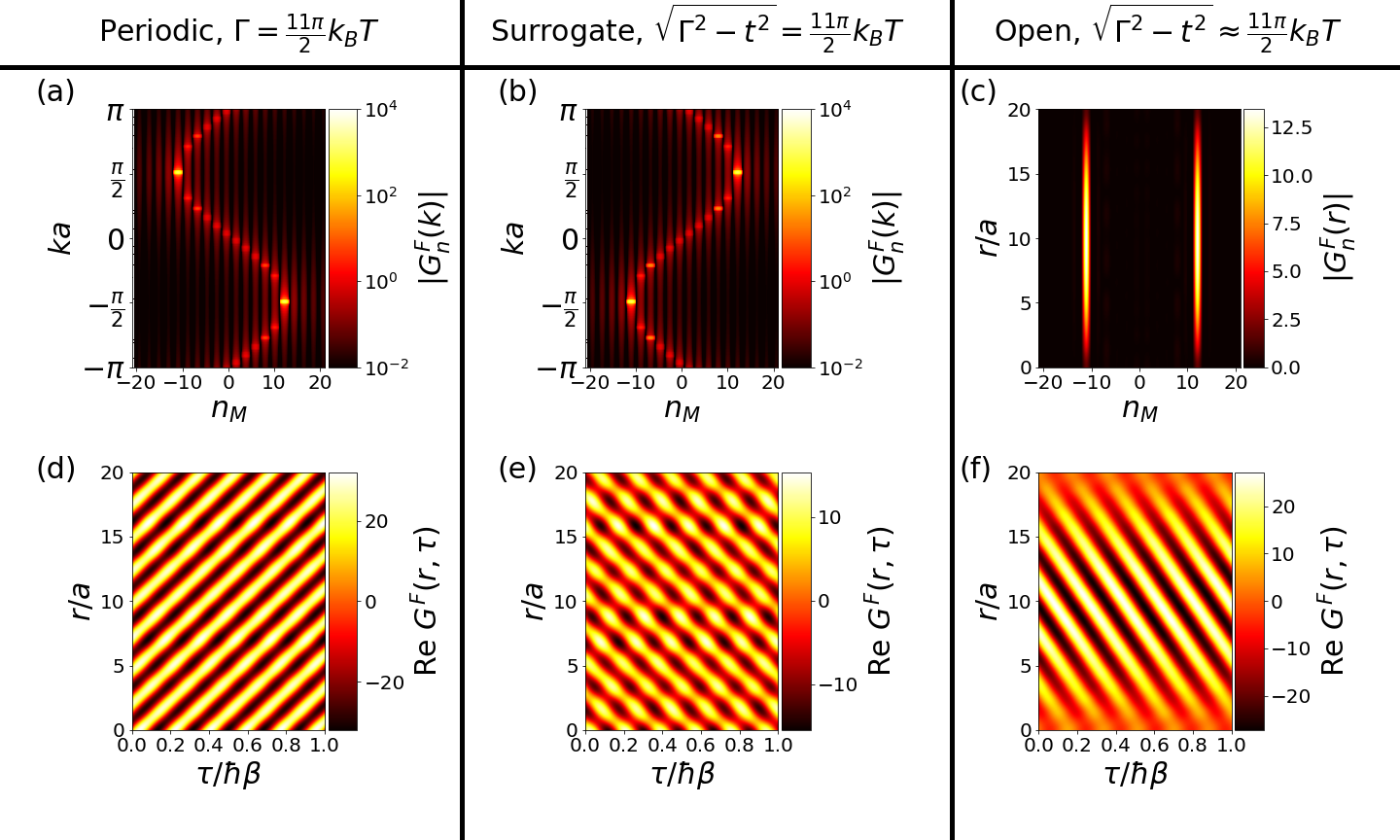}
	\caption{Green's function and its FT for the fermionic Hatano-Nelson model at resonance with $n_M=\pm 11$. Results are for PBC, surrogate Hamiltonian, and OBC for $\mu=-10^{-3}$, $t=0.1$ and $\beta=1$. We choose $\Gamma$ such that the resonance condition is met (see main text) for all these boundary conditions. A different choice of $\beta$ will only lead to a rescaling of $G$. We start by showing $|\widetilde{G}_n|$ as a function of the Matsubara mode $n_M$ and $k$ (distance to the left edge) for the periodic (open) system. A logscale is used for the periodic systems, such that the features besides the peaks of these functions are visible. This function is shown in (a) for the PBC system, in (b) for the surrogate Hamiltonian, and in (c) for the OBC system. Next, we show the real part of $G(r, \tau)$ as a function of imaginary time $\tau$ and distance $r$, where the spatial dependence is either computed using a FT in momentum, in the case of the periodic systems, or is computed from the wavefunctions, for the open system. The results are shown in (d) for PBC, in (e) for the surrogate Hamiltonian, and in (f) for the OBC system. The results for the imaginary part are similar. We use $100$ $k$-points for the periodic system and a lattice with $20$ sites for the open system.}
	\label{fig_G_Fermion_11pi}
\end{figure}
The FT of $G$, $\widetilde{G}_n$, is then given by 
\begin{equation}
	\widetilde{G}_n=\frac{1}{-i \hbar\omega_n-\mu+\mathbb{H}}=\sum\limits_{m} \phi_m^{R} {\phi_m^{L}}^\dagger\frac{1}{-i \hbar\omega_n+\zeta_m},
	\label{eq_G_n}
\end{equation}
where in the second equality we introduced twice the resolution of the identity $\sum_{m}\phi_m^R\phi_m^{L^\dagger}=\hat{1}$ and used the biorthogonality of the eigenvectors $\phi_m^{R/L}$ of $\mathbb{H}$ with eigenstate $\epsilon_m$. Notice that $\mathbb{H}$ is an operator in position/momentum and a matrix on the inner degrees of freedom of the fields, but it is not an operator in the Hilbert space. As such, its eigenvectors $\phi_m^{R,L}$ (which are basically the wavefunctions) are just vectors of functions of position/momentum. We see then that the behavior of $\widetilde{G}_n$, and consequently $G$, is determined by $\zeta_m$. In particular, the poles of this function are given by

\begin{equation}
	i\hbar\omega_n=\zeta_m,
	\label{eq_dis_point}
\end{equation}
which can only be satisfied (for real $T$ or $\beta$) when $\text{Re}{\zeta_m}=0$ and $\text{Im} \zeta_m=\hbar\omega_n=n_M  \pi \, k_BT$. Notice also that $\zeta/(k_BT)$ should be an integer multiple of $\pi$. Thus, this is possible only in the low (or intermediate) temperature limit.  When such conditions are met, both $G^B(\tau)$ and $G^F(\tau)$ take the form 
\begin{eqnarray}
	G\left(\tau\right)&\approx &\beta\sum\limits_{m} \phi_m^{R} {\phi_m^{L}}^\dagger \left(\text{Re}{\zeta_m}\right)^{-1} \, e^{-i\omega_{n}\tau},
	\label{eq_G_ress}
\end{eqnarray}
where $\omega_{n}$ and $\zeta_m$ almost satisfy Eq.~\ref{eq_dis_point}. A small deviation from this condition, encoded in a very small but finite $\text{Re}{\zeta_m}$, is necessary to prevent the function to diverge at the resonance. 

These systems then exhibit oscillatory behavior in imaginary time! Moreover, Eqs.~\eqref{eq_beta_res} are just solutions of Eq.~\eqref{eq_dis_point} for the Hatano-Nelson model, see Methods. Therefore, the oscillations in $\beta$ are indeed signatures of the iTC phase.	

It is remarkable that both $G$ and $G_n$ show a direct signature of the iTC phase. We present the results for this function close to the resonance for the fermion mode $n_M=\pm 11$ and $\mu=-10^{-3}$ in Fig.~\ref{fig_G_Fermion_11pi}. We plot $|\widetilde{G}_n^F|$ in Figs.~\ref{fig_G_Fermion_11pi} (a)-(c) in logscale to better reveal the presence of poles. For PBC [Fig.~\ref{fig_G_Fermion_11pi} (a)] and for the surrogate Hamiltonian [Fig.~\ref{fig_G_Fermion_11pi} (b)], we investigated it as a function of $k$, whereas for OBC [Fig.~\ref{fig_G_Fermion_11pi} (c)], we inspect $\widetilde{G}^F_n(r, 0)\equiv\widetilde{G}^F_n(r)$ as a function of a distance $r$ from the left edge of the system (position $0$ in our lattice). Starting from the PBC [Fig.~\ref{fig_G_Fermion_11pi} (a)], we observe the poles at $k=\pm \pi/2a$ and $n_M=\mp 11$ as expected, see Methods. The poles for different values of $n_M$ are much less intense as they follow a $-\sin(ka)$ function. They occur because we have set a small $\mu$ to prevent divergences in these functions. Similar features are seen for the surrogate Hamiltonian [Fig.~\ref{fig_G_Fermion_11pi} (b)], with the difference that now they follow a $\sin(ka)$ function, changing the sign relation between $n_M$ and $k$. The OBC does not allow for an analysis in momentum space, and we must take into account effects that are not present for the periodic systems. One is the NHSE, which localizes $\phi^R$ ($\phi^L$) in the right (left) edge of the system. The other is the modification of the spectrum due to the small size of the system, making the bandwidth equal to $2\sqrt{\Gamma^2-t^2}-\mathcal{O}(1/M)$. Hence, we need to look for conditions of resonance for this value of energy, instead of $2\sqrt{\Gamma^2-t^2}$.  However, considering such effects, the real space analysis [Fig.~\ref{fig_G_Fermion_11pi} (c)] reveals peaks precisely at the same values of $n_M$, showing that one can still clearly see this effect for a small system. 

The analysis of $\widetilde{G}_n$ explains directly the behavior of $G(\tau)$. Because there are poles in $k$, we also do a FT to obtain $G(\tau)$ in real space for PBC [Fig.~\ref{fig_G_Fermion_11pi} (d)] and the surrogate [Fig.~\ref{fig_G_Fermion_11pi} (e)], to directly compare with the OBC system [Fig.~\ref{fig_G_Fermion_11pi} (f)]. For PBC [Fig.~\ref{fig_G_Fermion_11pi} (d)], there are oscillations in both space and imaginary time. The period $\mathcal{T}= 2 /(11\hbar \beta)$ and wavelength $\lambda=4a$ are determined by $2\pi/\omega_n$ and $2\pi/k$. The behavior of this function is that of the plane wave in imaginary time $\exp\left\{i \left[r/\left(4a\right)-2\tau/\left(11\hbar \beta\right)\right]\right\}+c.c. $, so the bright lines observed in Fig.~\ref{fig_G_Fermion_11pi} (d)] are just wavefronts in imaginary time. For the surrogate [Fig.~\ref{fig_G_Fermion_11pi} (e)], similar patterns are visible but reversed, due to the fact that the poles in Fig.~\ref{fig_G_Fermion_11pi} (b) occur for opposite $n_M$ and $k$.
These features are more visible for OBC [Fig.~\ref{fig_G_Fermion_11pi} (f)], but with the complications commented before on the discussion for $\widetilde{G}_n$. Although there is a spatial decay due to the NHSE, the periodicity of this system does not change.

\begin{figure}[!ht]
	\centering
		\centering
		\includegraphics[width=\linewidth]{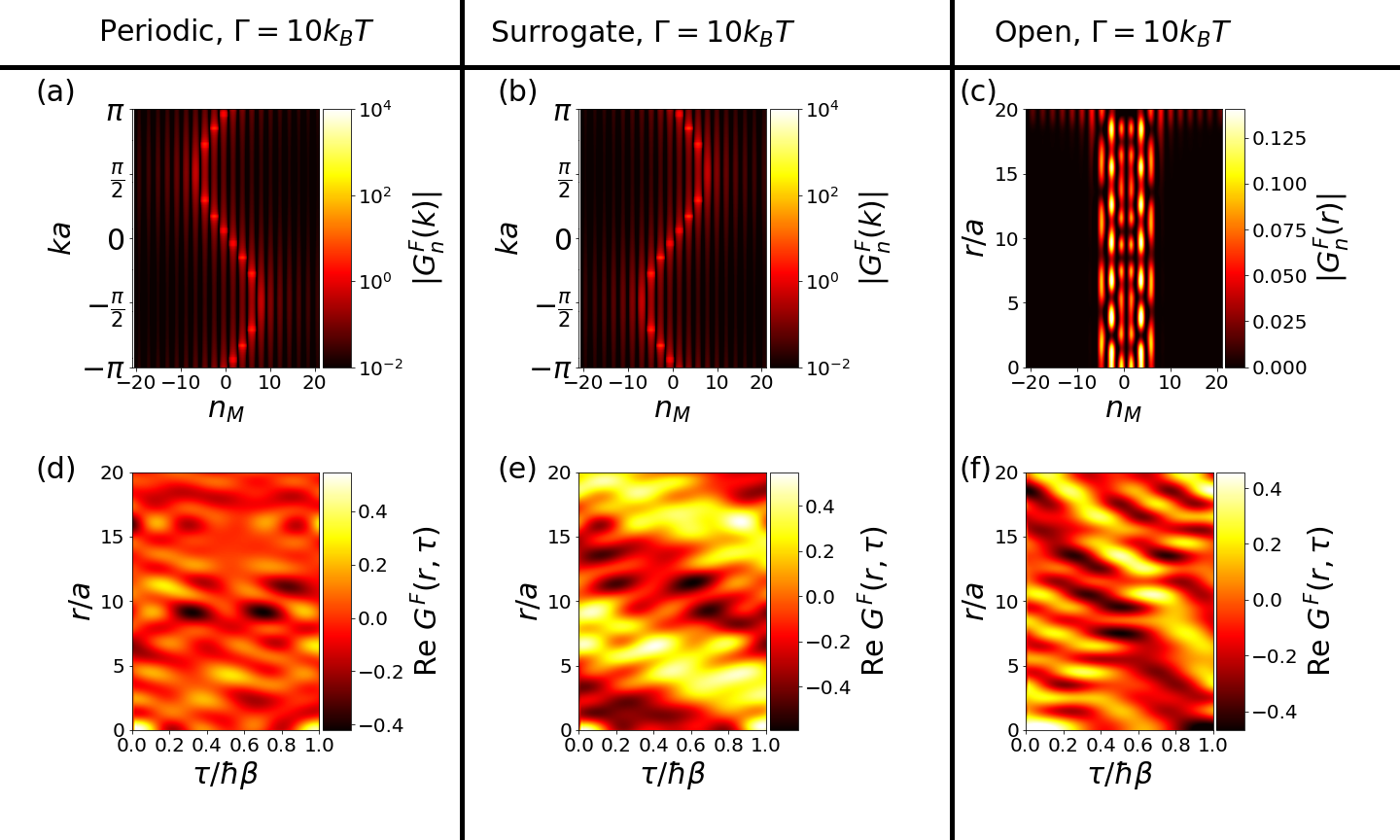}
		\centering
		\includegraphics[width=\linewidth]{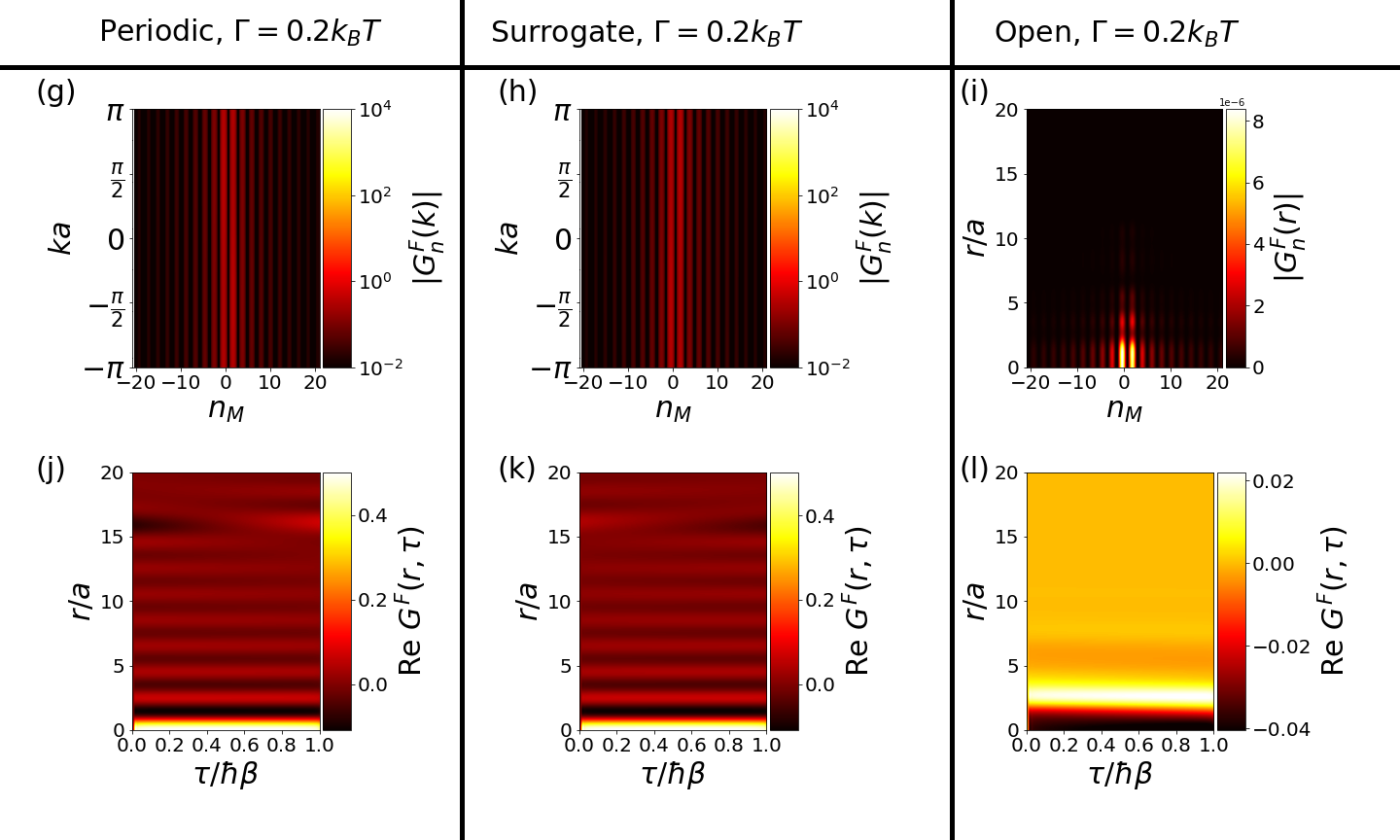}
	\caption{Green's function and its FT for the fermionic Hatano-Nelson model for non-resonant $\Gamma$. Results are for PBC, surrogate Hamiltonian, and OBC for $\mu=-10^{-3}$, $t=0.1$, and $\beta=1$. For $\Gamma=10 k_BT$ ($\beta \Gamma=10$), the results for $|\widetilde{G}_n|$ are shown in (a) for the PBC system, in (b) for the surrogate Hamiltonian, and in (c) for the OBC system. The real part of $G(r, \tau)$ is shown in (d) for PBC, in (e) for the surrogate Hamiltonian, and in (f) for the OBC system. For $\Gamma=0.2 k_BT$ ($\beta \Gamma=0.2$), the results for $|\widetilde{G}_n|$ are shown in (g) for the PBC system, in (h) for the surrogate Hamiltonian, and in (i) for the OBC system. The real part of $G(r, \tau)$ is shown in (j) for PBC, in (k) for the surrogate Hamiltonian, and in (l) for the OBC system. We use $100$ $k$-points for the periodic system and a lattice with $20$ sites for the open system.}
	\label{fig_G_Fermion_nres}
\end{figure}

If $\Gamma/(k_BT)$ is large, but not resonating, there will be much less intense poles at $\left|n_M\right|< 2\Gamma/\pi$. This is clearly seen for PBC in Fig.~\ref{fig_G_Fermion_nres} (a), where the resonances occur for $\left|n_M\right|=1, 3, 5$, but with a large spreading in momentum. Similar features are seen for the surrogate in Fig.~\ref{fig_G_Fermion_nres} (b). Such peaks are shown for OBC in Fig.~\ref{fig_G_Fermion_nres} (c), but with a stronger peak at $n_M=3$. Interestingly, the peaks in real space show different spatial periodicity, reflecting the fact that the peaks for different $n_M$ occur also for different $k$. This will influence the properties of $G(r, \tau)$. The absence of a sharp peak leads to an incoherent behavior, Fig.~\ref{fig_G_Fermion_nres} (d)-(f), resembling an amorphous phase. 

For small values of $\Gamma/(k_BT)$, as displayed in Fig.~\ref{fig_G_Fermion_nres} (g)-(l), there are no resonances at $n_M$ or $k$ for all boundary conditions [Fig.~\ref{fig_G_Fermion_nres} (g)-(i)]. Therefore, the system exhibits no oscillations in $\tau$ and only localization in $r$ Fig.~\ref{fig_G_Fermion_nres} (j)-(l)].

With these results, we understand which are the conditions for the occurrence of oscillations in imaginary time and real space. On resonance, Fig.~\ref{fig_G_Fermion_11pi}, there is ordering in both imaginary time and real space. For low temperatures and off resonance, Fig.~\ref{fig_G_Fermion_nres}, many peaks will be present simultaneously, blurring the oscillations in $\tau$. For high temperatures, Fig.~\ref{fig_G_Fermion_nres}, the oscillations are lost. The limit of $T\rightarrow 0$ ($\beta\rightarrow \infty$) in Eqs.~\eqref{eq_dis_point_PB} and \eqref{eq_dis_point_HN_Surr} implies that these conditions will be satisfied for $k=\pm \pi/2a$ and all values of $n_M$, such that the oscillations in $\tau$ will not be present. The behavior for the bosonic system is more intricate and is studied in the Supplementary Material.  Now, we discuss the meaning of these phases in more depth.

\clearpage
\section{Discussion}

Time crystals are phases where the continuous time translation symmetry is reduced to a discrete translation symmetry, in analogy to what happens for spatial translation invariance in a crystal \cite{shapere2012classical, wilczek2012quantum}. Even though the originally proposed model has some issues \cite{bruno2013impossibility, watanabe2015absence}, it generated a substantial offspring \cite{sacha2017time, khemani2019brief, sacha2020time}. From those, we highlight the recent proposal of time glasses \cite{verstraten2021time}, time quasicrystals \cite{autti2018observation} and dissipative time crystals \cite{lazarides2020time, booker2020non, kessler2020observation, sarkar2021signatures}, which are akin to the iTC phase. 

The later was briefly conjectured at the end of Ref.~\cite{wilczek2012quantum}, where the similarity between imaginary time and spatial dimensions in the Euclidean action are discussed. In analogy to the spatial variables, $\tau$ could also have preferred periods. In our case, Eq.~\eqref{eq_dis_point} sets this period to be $\mathcal{T}=2\pi /\omega_n=2\hbar \beta/(n_M)$ when  $n_M$ is in resonance with a value of $\zeta$. As the imaginary-time box has length $\hbar\beta$, for a resonance of Eq.~\eqref{eq_dis_point}, exactly $n_M/2$ oscillations will be in the imaginary time interval, analogously to a standing wave in space. This is precisely what is seen in Figs.~\ref{fig_G_Fermion_11pi} (d)-(f), where $11/2$ peaks or valleys are fitted in the imaginary time box. The same happens for bosonic systems, see Supplementary Material. 

The authors of Ref.~\cite{cai2020imaginary} studied this phase looking at a bosonic system coupled to a bath. They had a non-local action in imaginary time, which corresponds to a non-Markovian evolution of the system and leads to an (imaginary) time-dependent Hamiltonian. Nevertheless, the results found there are similar to ours. Their model presents a charge-density wave order and the order parameter related to this phase shows oscillations in both $\beta$ and $\tau$. Eqs.~\eqref{eq_dis_point} reveals that the peak for a specific Matsubara mode is also the peak for a specific $k$, see Methods. This will set a periodicity in space which, as seen from our results, survives also for a small OBC system, where momentum is no longer a good quantum number. 

The existence of these oscillations in space is not incidental. Eq.~\eqref{eq_dis_point} settles the condition for the presence of disorder lines for free systems \cite{timonin2021disorder} in the general theory of the Yang-Lee zeros. These phases were first obtained by Stephenson when studying the classical Ising model in a triangular lattice \cite{stephenson1970ising1}. The correlation function of the order parameter has an oscillatory part, together with the exponential decay, typical of critical systems. The characteristic modulation length follows scaling laws \cite{stephenson1970ising1, stephenson1970ising, stephenson1970range, stephenson1970two, chakrabarty2011modulation, chakrabarty2012universality} and is related to the presence of zeros of the partition function in the complex-parameter space. In the Euclidean action, $\tau$ is on the same footing as the spatial variables, so, intuitively, the presence of oscillations in imaginary time should not be surprising. The connection with disorder points also reveals that $\widetilde{G}_n$, for the value of $n$ that presents a resonance, is a natural order parameter. The characterization of disorder points is done by using the Fourier transform of the correlation function \cite{chakrabarty2011modulation, chakrabarty2012universality} and is analogous to the definition of the order parameter in charge density waves \cite{gruner1988dynamics}.  

However, imaginary time is distinct from real space in two aspects. First, bosons and fermions have different boundary conditions in $\tau$, leading to different Matsubara frequencies for each of them. Second, the imaginary-time interval is set by temperature, being proportional to $\beta$, and $\tau$ is conjugated to energy. Therefore, the evolution in imaginary time, given by $\exp(-H\tau)$, can be seen as a weighted projection on energy states \cite{de2011universal}. For $\tau\rightarrow \infty$ ($\beta\rightarrow\infty$), the system is projected to the ground state. The presence of a resonating condition implies that thare is a favored (imaginary) energy scale, which defines the period of oscillations in $\beta$. Due to pseudo-Hermiticity, the energies of the system come in complex conjugated pairs. Because the imaginary part of the energy is usually associated with dissipation, this can be interpreted as a kind of steady state, whereas the loss of probability in one mode is compensated by a gain in another. The fact that pseudo-Hermitian systems can have a biorthogonal unitary evolution \cite{gardas2016non} supports this view. Hence, such oscillations in temperature can be interpreted as a kind of resonating steady-state between system and reservoir, with the energy received/lost exhibiting peaks when the coupling of the system to the reservoir, given by $\Gamma$ in this case, resonates with a Matsubara frequency, which is a typical frequency associated to the thermal state.

In this work, we studied the thermodynamics of non-Hermitian quantum gases for finite temperatures, with special focus on the Hatano-Nelson model. The model was chosen because it exhibits phases with purely imaginary single-particle energies for both PBC and OBC. The presence of such modes lead to the iTC phase conjectured by Wilczek in Ref.~\cite{wilczek2012quantum}. The existence of this phase is revealed by both, the oscillation in the thermodynamic potentials as a function of $\beta$, and by the oscillations of the Green's function as a function of imaginary time. There is also an order in real space because the iTC is a disorder point. 

The iTC phase is interpreted as a resonance occurring precisely when the energy scale associated with the Matsubara frequencies matches the bandwidth of the imaginary part of the system, proportional to the coupling to the reservoir. Under this condition, the Matsubara frequency, usually interpreted as a mathematical tool, is manifested both in the thermodynamic and single-particle properties, and becomes measurable.  

The results obtained here for the Hatano-Nelson model should be present also in more complicated non-Hermitian models, as long as they exhibit modes with purely imaginary energies. Different models will lead to different crystalline structures in imaginary time. The presence of multiple orbitals and spin might unveil more intriguing phases. Studying these phases from the perspective of quantum heat engines and temperature dependent energy levels \cite{yunt2019topological} may also reveal new and interesting heat phenomena.

An important remark is that the fact that the system is non-Hermitian ultimately comes from the interaction of the system with a reservoir. Therefore, the introduction of imaginary terms in the energies comes from a (zero frequency) imaginary part of a self-energy, similarly to what was considered in Refs.~\cite{cai2020imaginary, cayao2022exceptional}. A frequency-dependent self-energy can lead to more intricate resonance conditions. A natural question concerning a more generic interacting system is the impact of many-body effects. In some discrete Floquet time-crystal, the effect of many-particle interactions is to stabilize the pre-thermal state, in which the time-crystalline behavior is observed, making it more robust \cite{else2020discrete}. Thermodynamic potentials are also multiparticle quantities and the effect of interaction can lead to combination of Matsubara frequencies in their observed behavior. Therefore, a further investigation of many-body effects in this description constitutes a promising topic for further research. 

As the Hatano-Nelson model can be engineered in many platforms, the iTC phase can be observed experimentally. The detection of these effects require their realization in a quantum platform. The measurement of a thermodynamic quantity will give direct demonstration of the iTC phase. Measurements of correlation functions for different times (further analytically continued to imaginary time) should also yield an indication of this phase. Moreover, evolution in imaginary time is related to response to quantum quenches \cite{de2011universal}. In this way, the response of the system to such a quench can also carry information on the periodicity in imaginary time. 

\section{Methods}

\subsection{Thermodynamic potentials of non-Hermitian quantum gases}\label{sec_thermo_gases}

For a non-Hermitian quantum gas, the energies may become complex and there are two kinds of eigenstates that label a microstate, $\ket{\left\{n_m\right\}}^R$ and $\ket{\left\{n_m\right\}}^L$, which are eigenstates of $H$ and $H^\dagger$, respectively, for each mode $m$ with energy $\epsilon_m$ ($\epsilon_m^*$) and eigenstate $\ket{m}^{R/L}$. The right eigenstates of $H$ alone do not form a complete set. However, the left and right eigenstates together do form a complete basis \cite{ashida2020non, moiseyev2011non, gardas2016non}. Using these states as a biorthogonal basis, one can compute the grand partition function and obtain the usual result for a quantum gas

\begin{eqnarray}
	\mathcal{Z}&=&\textrm{Tr}\left\{\exp\left[-\beta\left(\hat{H}-\mu\hat{N}\right)\right]\right\}\\
	&=&\sum\limits_{\left\{n\right\}}^{} \, ^L\braket{\left\{n_m\right\}|\left\{\exp\left[-\beta\left(\hat{H}-\mu\hat{N}\right)\right]\right\}|\left\{n_m\right\}}^R\nonumber\\
	&=&\prod\limits_{m}\sum\limits_{n_m}\exp\left[-\beta \, n_m\left(\zeta_m\right)\right],
\end{eqnarray}
where $\hat{N}$ is the number operator, $\mu$ is the chemical potential (we will consider it to be real), and we define for convenience $\zeta_m\equiv \epsilon_m-\mu$. 

The partition function is, in general, complex, but it can be real in some special cases. In the presence of a pseudo-Hermitian symmetry, $H$ and $H^\dagger$ are related by a similarity transformation $H^\dagger=g H g^{-1}, g=g^\dagger,$ such that their energies come in complex-conjugated pairs and their spectra are identical. As such, $Z$ (computed from $H$) and $Z^*$ (computed from $H^\dagger$) are the same, and consequently $Z$ is real \cite{gardas2016non}. In addition, parity-time (PT) symmetry makes the energies to be real, and in this case $Z$ is trivially real, as it is a sum of real numbers. In the following, it will become clearer that the most interesting effects for non-Hermitian gases will occur in the PT-symmetry broken phases, when the energies have a finite imaginary part, but the calculations presented here hold also if this symmetry is preserved.  

For bosons, $n_m\in \mathbb{N}$, whereas $n_m\in \left\{0,1\right\}$ for fermions. For bosons, we need to assume that $\text{Re} \, \epsilon\geq \mu $ ($\text{Re}\, {\zeta_m}\geq 0$), $\left|\exp\left[-\beta \, n_m \zeta_m\right]\right|\leq 1$, otherwise this Laurent series diverges. The sum yields the familiar \cite{salinas2001introduction,huang1963statistical} partition functions for bosons and fermions,
\begin{eqnarray}
	\mathcal{Z}_\text{B/F}&=&\prod\limits_{m}\left[1\mp\exp\left(-\beta \zeta_m\right)\right]^{\mp 1}.
	\label{eq_Z_B}
\end{eqnarray}
From $\mathcal{Z}$, one can obtain all the thermodynamic quantities. Some thermodynamic potentials are particularly interesting to consider and are discussed in the Methods.

The first is the average occupancy of each level:
\begin{equation}
	\braket{\hat{N}_m}=\frac{1}{\beta}\frac{\partial\ln\left(\mathcal{Z}\right)}{\partial \mu}, 
\end{equation}
which leads to the Bose-Einstein and Fermi-Dirac distributions,
\begin{equation}
	g_{\text{B/F}}\left(\beta, \mu, \zeta_m\right)=\frac{1}{\beta}\frac{\partial\ln\left(\mathcal{Z}_{\text{B/F}}\right)}{\partial \mu}=\sum\limits_{m}\left[\exp\left(\beta \zeta_m\right)\mp 1\right]^{-1}.
	\label{eq_f_B_F}
\end{equation}
Other kinds of potentials that are interesting are the ones related to the thermal behavior of the system. These are the grand canonical potential $F=-(1/\beta)\ln\mathcal{Z}$,
\begin{equation}
	F_{\text{B/F}}=\pm \frac{1}{\beta}\sum\limits_{m}\ln\left[1\mp\exp\left(-\beta \zeta_m\right)\right],
	\label{eq_Omega_B_F}
\end{equation}
the internal energy $U$
\begin{equation}
	U_{\text{B/F}}=\sum\limits_{m}\epsilon_m g_{\text{B/F}}\left(\beta, \mu, \zeta_m\right),
	\label{eq_U_B_F}
\end{equation}
and the entropy $S=-\partial F/\partial T=k_B\beta^2\partial F/\partial \beta$, which reads 
\begin{eqnarray}
\hspace{-1cm}	\frac{S_{\text{B/F}}}{k_B}&=&-\beta F_{\text{B/F}}+\beta\sum\limits_{m}\zeta_m\beta  g_{\text{B/F}}\left(\beta, \mu, \zeta_m\right)=-\beta F_{\text{B/F}}+\beta U_{\text{B/F}}-\beta \mu N,
	\label{eq_S_B_F}
\end{eqnarray}
where $N=\sum\limits_{m}\braket{\hat{N}_m}$ is the total number of particles. Notice that this just follows from the thermodynamic definition of $F$ \cite{salinas2001introduction}.

These systems have a classical behavior in the limit $\left|\exp(-\beta \mu)\right|\ll 1$ \cite{salinas2001introduction}, when the length scale of thermal fluctuations (proportional to $1/T$) are smaller than the quantum ones (proportional to $\mu$), which is related to high temperatures or small densities. In this situation, both functions reduce to the Boltzmann distribution
\begin{equation}
	g_{F/B}(\beta, \mu)\rightarrow g_{cl}=\frac{1}{\mathcal{Z}_{cl}}\sum\limits_{m}\exp(-\beta  \zeta_m).
	\label{eq_f_cl}
\end{equation}

The thermodynamic potentials then assume the form
\begin{eqnarray}
	F_{cl}(\beta, \mu)=-\frac{1}{\beta}\sum\limits_{m}\exp\left(-\beta \zeta_m\right)\label{eq_Omega_cl},\\
	U_{cl}(\beta, \mu)=\sum\limits_{m}\epsilon_m\exp\left(-\beta \zeta_m\right)\label{eq_U_cl},\\
	S_{cl}(\beta, \mu)/k_B=\sum\limits_{m}\left(1+\beta \zeta_m\right)\exp\left(-\beta \zeta_m\right)\label{eq_S_cl}.
\end{eqnarray}

\subsection{Poles of the Hatano-Nelson model}

For the system with OBC, one needs to numerically diagonalize $\mathbb{H}$, to obtain the spectrum and the wavefunctions. As the Hatano-Nelson model has only one site per unit cell, for the periodic and surrogate Hamiltonians $\mathbb{H}$ will be just $\epsilon(k)$ and $\epsilon_{\text{surr}}(k)$, respectively, which are numbers, and consequently do not have eigenvectors. The PBC system has the spectrum 
\begin{equation}
	\epsilon(k)=-2 t \cos(ka)- 2i\Gamma \sin(ka),
\end{equation}
where $a$ is the lattice parameter. If we just replace this $\epsilon(k)-\mu$ in Eq.~\eqref{eq_dis_point}, we obtain the condition for resonance
\begin{equation}
	i\hbar\omega_n=-2t \cos(ka)-2i \Gamma\sin(ka)-\mu,
\end{equation}
which is satisfied for
\begin{equation}
	k=\frac{1}{a}\arccos\left(-\frac{\mu}{2t}\right), \qquad\Gamma\sin(ka)=-n_M\frac{\pi}{2} k_BT.
	\label{eq_dis_point_PB}
\end{equation} 
Bosonic systems should satisfy $\mu\leq-2\left|t\right|$. Therefore, for them the only possible solution for the above equations is $k=0$ and $n_M=0$, which occurs when $\mu=-2t$. Conversely, for fermions $\mu$ is not restrict and if we choose $\mu=0$, the resonance condition simplifies to $k=\pm \pi/(2a)$ and $\Gamma=\mp n_M (\pi/2) k_BT$.

For OBC, a simple analysis is not really feasible. Hence, we turn to the surrogate Hamiltonian, as it has the same bulk spectrum. Using the band dispersion
\begin{equation}
	\epsilon_{\text{surr}}(k)=\sqrt{\left|\Gamma^2-t^2\right|}\left[\text{sgn}\left(\Gamma-t\right)e^{ika}-\text{sgn}\left(\Gamma+t\right)e^{-ika}\right],
\end{equation}
the condition for resonance becomes
\begin{equation}
	\hspace{-0.17cm}i\hbar\omega_n=\begin{cases}2i\, \text{sgn}\left(\Gamma\right)\sqrt{\left|\Gamma^2-t^2\right|}\sin(ka)-\mu,\quad \left|\Gamma\right|>\left|t\right|,\\-2\,\text{sgn}\left(t\right)\sqrt{\left|\Gamma^2-t^2\right|}\cos(ka)-\mu,\quad \left|\Gamma\right|<\left|t\right|,\end{cases}
\end{equation}
and we find the solutions
\begin{equation}
	\begin{matrix}
		\mu=0,&\sqrt{\Gamma^2-t^2}\sin(ka)=\text{sgn}(\Gamma)n_M\frac{\pi}{2} k_BT,&\left|\Gamma\right|>\left|t\right|,\\
		k=\arccos\left(-\text{sgn}(t)\frac{\mu}{2\sqrt{t^2-\Gamma^2}}\right), &n_M=0,&\left|\Gamma\right|<\left|t\right|.
	\end{matrix}
	\label{eq_dis_point_HN_Surr}
\end{equation}

In the PT-broken phase, $\left|\Gamma\right|>\left|t\right|$, and the first of Eqs.~\eqref{eq_dis_point_HN_Surr} can be rewritten as
\begin{equation}
	k=\frac{1}{a}\text{sgn}(\Gamma)\arcsin\left[- \frac{n_M}{2\sqrt{\Gamma^2-t^2}/\left(\pi k_BT\right)}\right],
\end{equation}
such that there will be poles at all values of $n_M$ that are smaller than the ratio $2\sqrt{\Gamma^2-t^2}/\left(\pi k_BT\right)$. For a finite system, however, $k$ is of the form $k=n\pi /(M a), -M< n\leq M$, such that some of these modes can be present only for an infinite lattice, where $k$ can take any value between $-\pi/a$ and $\pi/a$.


\subsection*{Supplementary information}
The Supplemental Methods contains details about the Hatano-Nelson model, discusses second quantization and path integrals for non-Hermitian operators and show the results for the Green's functions for bosonic systems.

\subsection*{Acknowledgments}
We thank F. Wilczek, T. Hans Hansson, M. Bourennane, B. Buca, and A. Moustaj  for very interesting discussions about our results. We specially thank V. Gritsev for pointing out the connection of our results with the theory of Yang-Lee-Fisher zeros and Fernando Nicacio and Lumen Eek for noticing some typos in the first version of the manuscript. 

\subsection*{Funding}

RA acknowledges funding from the Brazilian Coordination for the Improvement of Higher Education Personnel (CAPES), from the Brazilian National Council for Research and Development (CNPq), from the Delta Institute for Theoretical Physics (DITP) consortium, a program of the Netherlands Organization for Scientific Research (NWO), and by the Knut and Alice Wallenberg Foundation through the Wallenberg Academy Fellows program. ECM was supported in part by CNPq, FAPERJ and CAPES.

\subsection*{Data availability}

All data analyzed during the current study is available on reasonable request. 

\subsection*{Author contributions}
R.A. did the calculations under the supervision of C.M.S and E.C.M. All authors contributed to the interpretation of the results and to the manuscript. 

\subsection*{Competing interests}
The authors declare no competing interests.

\printbibliography

\clearpage

\renewcommand{\figurename}{Supplementary Figure}
\section{Supplementary Methods}\label{sec_Hatano-Nelson}

\subsection{The Hatano-Nelson model}
\subsubsection{Spectrum, dispersion relation and phases}

In this Section, we discuss in detail the Hatano-Nelson model. It is one of the earliest examples of non-Hermitian models \cite{hatano1996localization} and is the simplest one to realize non-Hermitian topological phases \cite{bergholtz2019exceptional}. In second quantization, its Hamiltonian reads    
\begin{equation}
	H=-\sum\limits_{i}^M \left[\left(t-\Gamma\right) c_j^\dagger c_{j+1}+\left(t+\Gamma\right) c_{j+1}^\dagger c_{j}\right],
\end{equation}
where $M$ is the lattice size and $c_j$ ($c_j^\dagger$) annihilates (creates) a particle at site $j$. This is a simple hopping model in $1D$ with a reciprocal part (proportional to $t$) and a nonreciprocal part (proportional to $\Gamma$).

For non-Hermitian systems, the spectrum can be different for different boundary conditions. This is also the case for the Hatano-Nelson model. This occurs basically because of the breaking of Bloch theorem, as the wavefunctions are localized due to the non-Hermitian skin effect. One can then recover the bulk spectrum by doing an analytical continuation of the momentum, i. e., introducing an imaginary part that accounts for this localization. The Hamiltonian obtained then, the so called surrogate Hamiltonian \cite{yao2018edge, gong2018topological, lee2019unraveling}, describes very well the bulk bands obtained for OBC. 

In the case of the Hatano-Nelson model, the band dispersion for PBC is
\begin{equation}
	\epsilon(k)=-2t\cos(k a)-2i\Gamma\sin(k a),
	\label{eq_disp_Hatano-Nelson}
\end{equation}
where $a$ is the lattice parameter, while for the surrogate Hamiltonian it is (see Section \ref{app_surr}) 
\begin{eqnarray}
	\hspace{-0.7cm}\epsilon_{\text{surr}}(k)&=&\begin{cases}2i\,\text{sgn}(\Gamma)\sqrt{\left|\Gamma^2-t^2\right|}  \sin(ka), \left|\Gamma\right|>\left|t\right|,\\-2\,\text{sgn}(t)\sqrt{\left|\Gamma^2-t^2\right|}  \cos(ka), \left|\Gamma\right|<\left|t\right|.\end{cases}
	\label{eq_disp_surr_1}
\end{eqnarray}

\begin{figure}[!h]
	\centering
	\includegraphics[width=\linewidth]{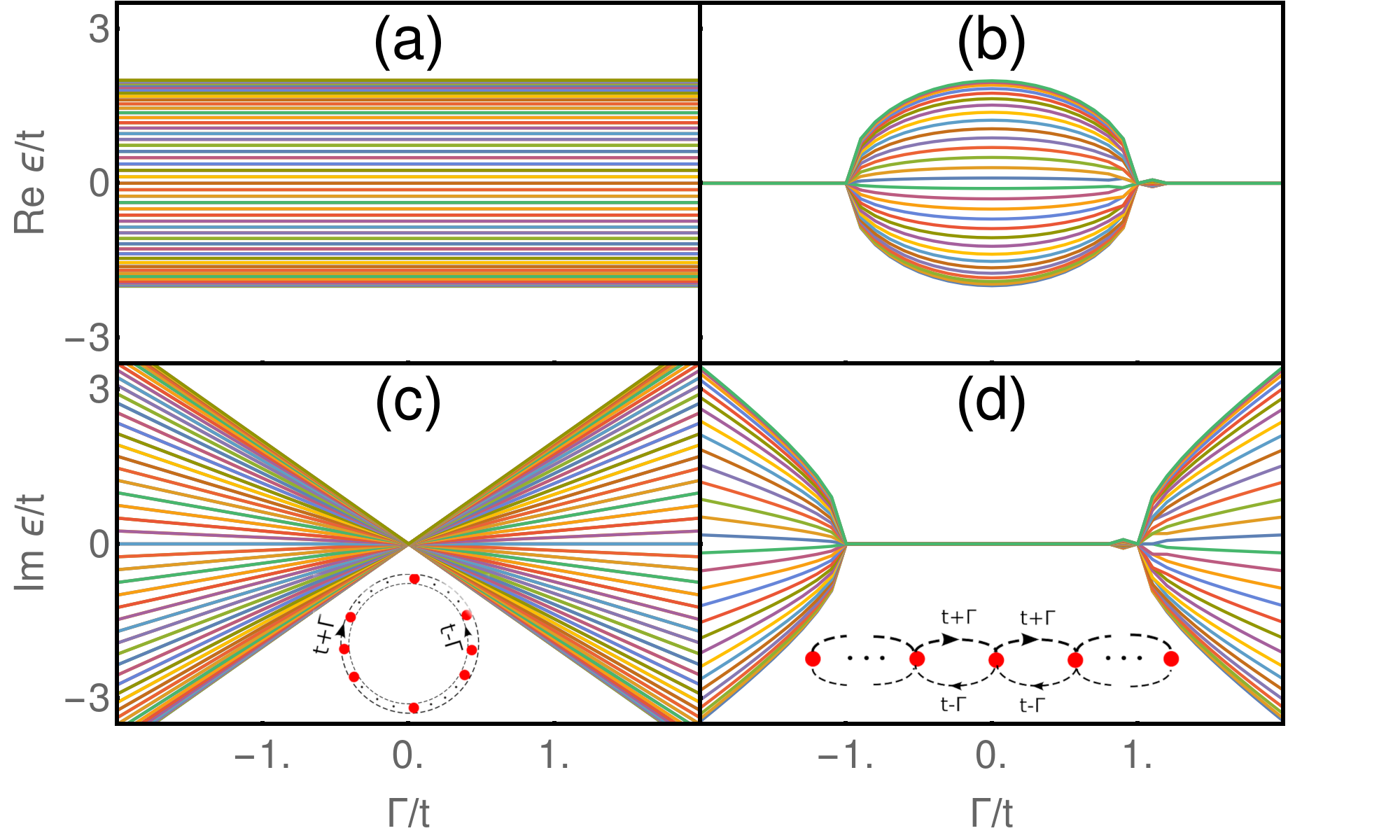}
	\caption{Spectrum of the Hatano-Nelson model for PBC and OBC as a function of $\Gamma/t$. The real part of the spectrum is shown in (a) for PBC and in (b) for OBC. The imaginary part is shown in (c) for PBC and in (d) for OBC. The spectrum is calculated for $100$ $k$-points, corresponding to a lattice of $100$ sites for the periodic system, and $20$ sites for the open system. Inset: representations of the lattice for (c) PBC and (d) OBC.}
	\label{fig_spec}
\end{figure}	

In Supplementary Figure~\ref{fig_spec}, we show the spectrum of this system for both boundary conditions. For PBC, the real part of the spectrum is exactly the spectrum of a $1D$ tight-binding model, with a continuum of bands going from $-2t$ to $2t$. The system with OBC displays a more interesting behavior. For $\left|\Gamma/t\right|<1$, it resembles also the spectrum of the simple hopping model, but with a bandwidth of $4\sqrt{\Gamma^2-t^2}$ instead of $4t$. More surprisingly, for $\left|\Gamma/t\right|>1$, the real part of the energy vanishes. Thus, for the system with OBC, the non-reciprocity caused by the coupling to the reservoir changes significantly even the real part of the spectrum. 

The imaginary part of the spectrum is however the most interesting one in a non-Hermitian system. For the PBC system, there is again a continuum of bands, but now following $\sin(ka)$ instead of a $\cos(ka)$ and with a bandwidth directly proportional to $\Gamma$. Once again, the system with OBC shows more unconventional features, as there are finite imaginary values of the energy only for $\left|\Gamma/t\right|>1$, when the energy is completely imaginary! The bandwidth is proportional to $\sqrt{\Gamma^2-t^2}$, growing almost linearly with $\Gamma$ for large $\Gamma$. Inspection of the spectrum shows an accumulation of modes at $\Gamma=\pm t$, signaling the non-Bloch band collapse \cite{lee2019unraveling, alvarez2018non, longhi2020non, longhi2020_nonbloch, arouca2020unconventional}. At this point, the imaginary part of the momentum diverges (see Section \ref{app_surr}), parity-time (PT) symmetry is broken, and the system is only pseudo-Hermitian, with the energy coming in complex conjugated pairs.   

For its simplicity and because it can show arbitrarily large imaginary parts in its spectrum (for both boundary conditions), the Hatano-Nelson is our model of choice to showcase the unique thermal features of non-Hermitian systems. 

\subsubsection{Surrogate Hamiltonian}\label{app_surr}

From the second quantized version of the Hatano-Nelson model one can readily obtain this Hamiltonian in first quantization
\begin{equation}
	H=-\sum\limits_{j=1}^M \left[\left(t-\Gamma\right) \ket{j} \bra{ j+1}+\left(t+\Gamma\right) \ket{j+1}\bra{j}\right],
\end{equation}
where $\ket{j}$ denotes a one-particle state localized on site $j$.

Then, the (time-independent) Schrödinger equation for a eigenstate $\ket{\psi_\epsilon}^R$ with energy $\epsilon$ takes the form
\begin{eqnarray}
	H\ket{\psi_\epsilon}^R&=&-\sum\limits_{j=1}^M \left[\left(t-\Gamma\right) \phi^R_\epsilon(j+1)\ket{j}+\left(t+\Gamma\right) \phi^R_\epsilon(j)\ket{j+1}\right]=\epsilon\ket{\psi_E}^R \nonumber\\\therefore \braket{l| H|\psi_\epsilon}^R&=& \left(\Gamma-t\right) \phi^R_\epsilon(l+1)-\left(\Gamma+t\right) \phi^R_\epsilon(l-1)=\epsilon\phi^R_\epsilon(l),
\end{eqnarray}
where we used that the wavefunction on site $j$ is $\braket{ j|\psi_\epsilon}^R=\phi^R_\epsilon(j)$ and $\braket{ j|l}=\delta_{j,l}$.

One can now assume that the wavefunction has a non-Bloch form \cite{lee2019anatomy, yao2018edge, lee2019unraveling}
\begin{equation}
	\phi^R_\epsilon(l)= e^{i \left[k(\epsilon)+i\kappa(\epsilon)\right](l-j) a}\phi^R_\epsilon(j),
\end{equation}
with an imaginary part $\kappa$ of the momentum, which will take into account the localization of the wavefunction. 

Then, one can convert the Schrödinger equation in an equation for $k$, $\kappa$ and $\epsilon$,
\begin{eqnarray}
	\epsilon&=&\left(\Gamma-t\right) e^{i \left[k(\epsilon)+i\kappa(\epsilon)\right] a}-\left(\Gamma+t\right) e^{-i \left[k(\epsilon)+i\kappa(\epsilon)\right] a}.
\end{eqnarray}

Assuming that $\kappa$ does not depend on $\epsilon$, but only on the parameters of the model, this equation simplifies to the dispersion relation of the surrogate Hamiltonian
\begin{eqnarray}
	\epsilon_{\text{surr}}(k)&=&\left(\Gamma-t\right) e^{i \left[k(\epsilon)+i\kappa\right] a}-\left(\Gamma+t\right) e^{-i \left[k(\epsilon)+i\kappa\right] a}.
	\label{eq_disp_surr_1_app}
\end{eqnarray}

We can choose a specific value of $\epsilon$ to obtain $\kappa$. Choosing $\epsilon=0$
\begin{eqnarray}
	0&=&\left(\Gamma-t\right) e^{i \left[k(0)+i\kappa\right] a}-\left(\Gamma+t\right) e^{-i \left[k(0)+i\kappa\right] a}\nonumber\\
	\therefore \left(\Gamma-t\right) e^{i \left[k(0)+i\kappa\right] a}&=&\left(\Gamma+t\right) e^{-i \left[k(0)+i\kappa\right] a}\nonumber,\\
	\therefore  e^{2i \left[k(0)+i\kappa\right] a}&=&\frac{\left(\Gamma+t\right)}{\left(\Gamma-t\right)} \rightarrow e^{-\kappa a}=\sqrt{\left|\frac{\left(\Gamma+t\right)}{\left(\Gamma-t\right)}\right|},\nonumber\\
	 \therefore\kappa&=&\frac{1}{a}\log\left(\sqrt{\left|\frac{\left(\Gamma-t\right)}{\left(\Gamma+t\right)}\right|}\right) .
\end{eqnarray}

Substituting then the expression of $\exp(-\kappa a)$ in Eq.~\eqref{eq_disp_surr_1_app}, one obtains 
\begin{eqnarray}
	\epsilon_{\text{surr}}(k)&=&\left(\Gamma-t\right)e^{-\kappa a} e^{i k a}-\left(\Gamma+t\right)e^{\kappa a} e^{-i k a}\nonumber\\
	&=&\left(\Gamma-t\right)\sqrt{\left|\frac{\left(\Gamma+t\right)}{\left(\Gamma-t\right)}\right|} e^{i k a}-\left(\Gamma+t\right)\sqrt{\left|\frac{\left(\Gamma-t\right)}{\left(\Gamma+t\right)}\right|} e^{-i k a}\nonumber\\
	&=&\sqrt{\left|\Gamma^2-t^2\right|}\left[\text{sgn}\left(\Gamma-t\right) e^{i k a}-\text{sgn}\left(\Gamma+t\right) e^{-i k a}\right].
	\label{eq_disp_surr_2_app}
\end{eqnarray}

If $t>0$ and $\Gamma>t$, $$\text{sgn}\left(\Gamma-t\right)=\text{sgn}\left(\Gamma+t\right)=1$$ and $\epsilon_{\text{surr}}=2 i\sqrt{\left|\Gamma^2-t^2\right|}\sin(ka)$. If $t>0$ and $\Gamma< -t$ $$\text{sgn}\left(\Gamma-t\right)=\text{sgn}\left(\Gamma+t\right)=-1$$ and $\epsilon_{\text{surr}}=-2 i\sqrt{\left|\Gamma^2-t^2\right|}\sin(ka)$. However, if $t>0$ and $-t<\Gamma<t$, $$\text{sgn}\left(\Gamma-t\right)=-\text{sgn}\left(\Gamma+t\right)=-1$$ and $\epsilon_{\text{surr}}=-2 \sqrt{\left|\Gamma^2-t^2\right|}\cos(ka)$. For negative $t$, a similar argument holds, but with a difference in sign in the last equality. We can then write the expression for $\epsilon_{\text{surr}}(k)$ as 
\begin{eqnarray}
	\hspace{-0.7cm}\epsilon_{\text{surr}}(k)&=&\begin{cases}2i\text{sgn}(\Gamma)\sqrt{\left|\Gamma^2-t^2\right|}  \sin(ka), \left|\Gamma/t\right|>1,\\-2\text{sgn}(t)\sqrt{\left|\Gamma^2-t^2\right|}  \cos(ka), \left|\Gamma/t\right|<1,\end{cases}
	\label{eq_disp_surr_3_app}
\end{eqnarray}
which is precisely Eq.~\eqref{eq_disp_surr_1} of the main text. The spectrum of such model, shown in Supplementary Figure~\ref{fig_spec_surr}, is identical to the one for open boundary conditions displayed in Supplementary Figure~\ref{fig_spec}.

\begin{figure}[!h]
	\centering
	\includegraphics[width=0.8\linewidth]{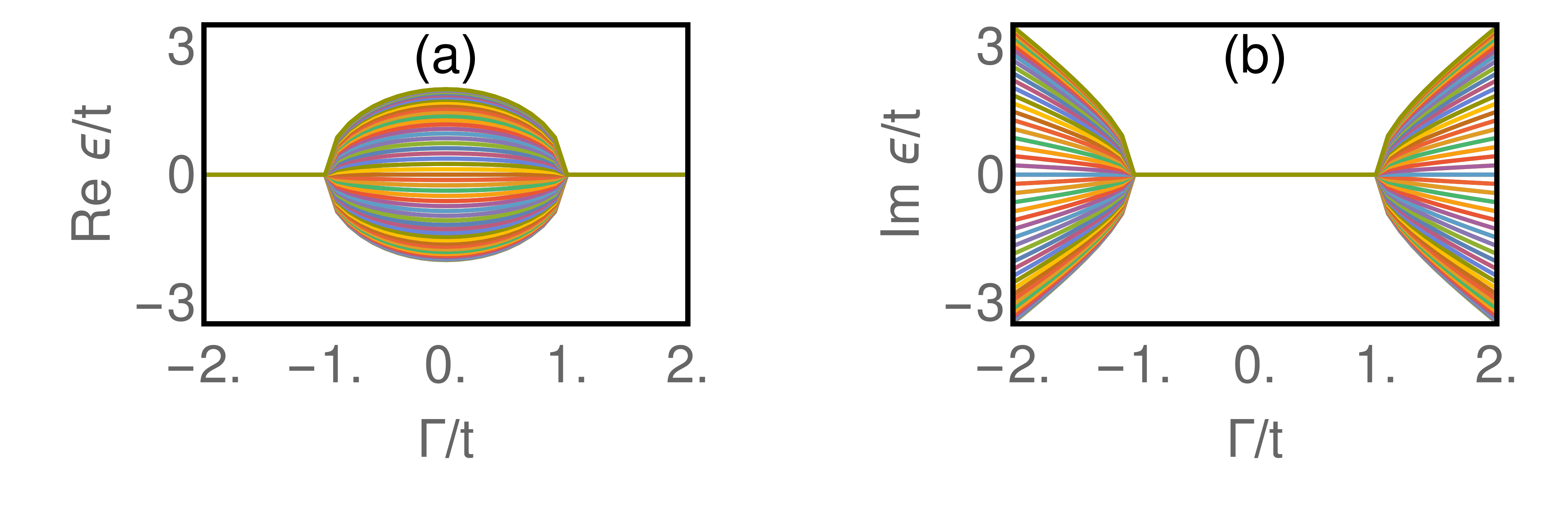}
	\caption{Real and imaginary spectrum of the surrogate Hamiltonian of the Hatano-Nelson as a function of $\Gamma/t$. (a) The real part of the spectrum; (b) the imaginary part. The spectrum is calculated for $100$ $k$-points, corresponding to a lattice of $100$ sites.}
	\label{fig_spec_surr}
\end{figure}

\section{Biorthogonal second quantization and coherent states}\label{app_nh_second}
Let us define the eigenstate $\ket{m}^R$ as being given by a creation operator $R_m^\dagger$ acting on the vacuum state $\ket{0}^R$,
\begin{equation}
	\ket{m}^R=R_m^\dagger\ket{0}^R,
\end{equation}
and a similar relation holds between $\ket{m}^L$ and a creation operator $L^\dagger_m$,
\begin{equation}
	\ket{m}^L=L_m^\dagger\ket{0}^L.
\end{equation}

These eigenstates must satisfy the biorthogonal relations. A very natural way to do it, is to impose (anti)comutation relations \cite{bandyopadhyay2020classification, chen2021entanglement} between $R^\dagger$ and $L$
\begin{equation}
	^L\hspace{-0.1cm}\braket{m|l}^R=^L\hspace{-0.1cm}\braket{0|L_m R^\dagger_l|0}^R=^L\hspace{-0.1cm}\bra{0}\pm R^\dagger_lL_m+\left[L_m,R^\dagger_l\right]_{\mp}\ket{0}^R=\delta_{m,l},
\end{equation}
where we use that $^L\hspace{-0.1cm}\braket{0|0}^R=1$, that $L_m$ annihilates $\ket{0}^R$, $L_m\ket{0}^R=0$, and that the commutator $\left[L_m,R^\dagger_l\right]_{-}$ $\left(\textrm{anti-commutator} \left[L_m,R^\dagger_l\right]_{+}\right)$ satisfies $\left[L_m,R^\dagger_l\right]_{\mp}=\delta_{m,l}$, with the upper sign for bosons and the lower one for fermions.

Using these relations, the Hamiltonian reads
\begin{equation}
	H=\sum\limits_{m}\epsilon_m \ket{m}^{R\, L}\hspace{-0.1cm}\bra{m}=\sum\limits_{m}\epsilon_m R_m^\dagger|{0}^{R\, L}\hspace{-0.1cm}\bra{0}L_m,
\end{equation}
such that 
\begin{equation}
	H\ket{l}^R=\sum\limits_{m}\epsilon_m R_m^\dagger\ket{0}^{R\, L}\hspace{-0.1cm}\bra{0}L_mR^\dagger_l\ket{0}^R=\epsilon_l R_l^\dagger\ket{0}^{R}=\epsilon_l \ket{l}^R,
\end{equation}
and 
\begin{equation}
	^L\hspace{-0.1cm}\bra{l}H=\sum\limits_{m}\epsilon_m \, ^L\hspace{-0.1cm}\bra{0}L_lR_m^\dagger\ket{0}^{R\, L}\hspace{-0.1cm}\bra{0}L_m=\epsilon_l \,^{ L}\hspace{-0.1cm}\bra{0}L_l=\epsilon_l\, ^{ L}\hspace{-0.1cm}\bra{l}.
\end{equation}
Similar relations hold between $R$ and $L^\dagger$, 
\begin{eqnarray}
	&&R_m\ket{0}^L=0,\qquad \left[R_m,L^\dagger_l\right]_{\mp}=\delta_{m,l},
	\nonumber\\ &&H^\dagger=\sum\limits_{m}\epsilon_m^* \ket{m}^{L\, R}\hspace{-0.1cm}\bra{m}=\sum\limits_{m}\epsilon_m^* L_m^\dagger\ket{0}^{L\, R}\hspace{-0.1cm}\bra{0}R_m.
\end{eqnarray}

In addition, we have that
\begin{equation}
	\hspace{-0.5cm}\left[R_m, R_l\right]_\mp=\left[L_m, L_l\right]_\mp=\left[L_m^\dagger, L_l^\dagger\right]_\mp=\left[R_m^\dagger, R_l^\dagger\right]_\mp=0,
\end{equation}
such that the action of $R^\dagger$ and $L$ ($L^\dagger$ and $R$) on many-body right (left) states is the same as regular creation and annihilation operators \cite{sakurai2014modern, stoof2009ultracold}. In particular, the eigenvalues of the right (left) number operator $N^R_m$ ($N^L_m$) of a mode $m$, defined as $R_m^\dagger L_m$ ($L_m^\dagger R_m$), are the natural numbers. For fermions, only $0$ and $1$.

Using these properties, we can define coherent states \cite{stoof2009ultracold} of a given mode
\begin{equation}
	\ket{\psi_m}_m^R=e^{\pm \psi_m R^\dagger_m}\ket{0}^R\qquad \ket{\psi_m}_m^L=e^{\pm \psi_m L^\dagger_m}\ket{0}^L,
\end{equation}
with the upper sign for bosons and the lower one for fermions, which are eigenstates of $L_m$ and $R_m$, 
\begin{equation}
	L_m\ket{\psi_m}_m^R=\psi_m \ket{\psi_m}_m^R,\qquad R_m\ket{\psi_m}_m^L=\psi_m \ket{\psi_m}_m^L.
\end{equation}  
Note that for fermions, $\psi_m$ is a Grassmann variable.

We can then define a coherent state for all modes
\begin{equation}
	\ket{\Psi}^R=e^{\pm \sum\limits_m\psi_m R^\dagger_m}\ket{0}^R,\qquad \ket{\Psi}^L=e^{\pm \sum\limits_{m}\psi_m L^\dagger_m}\ket{0}^L,
\end{equation} 
where $\Psi=\left(\begin{matrix}\psi_0&\psi_1&\dots\end{matrix}\right)$ is a vector of the $\psi_m$ parameters.
The coherent states are not biorthogonal
\begin{equation}
	^{L}\hspace{-0.1cm}\braket{\Psi|\Psi'}^{R}=e^{\pm \Psi^\dagger \Psi'},
\end{equation}
but they do form a supercomplete set 
\begin{equation}
	\int d\Psi^\dagger d\Psi \,  e^{- \Psi^\dagger \Psi}  \ket{\Psi}^{R \, L} \hspace{-0.1cm}\bra{\Psi}=\hat{1},
\end{equation}
and we can use them as a basis to write any operator.

In particular, 
\begin{align}
	\begin{split}
	\textrm{Tr}\left[\hat{O}\right]&=\sum\limits_{\left\{n_m\right\}} \hspace{-0.1cm}^L\braket{n_m|\hat{O}|n_m}^R\\
		&=\sum\limits_{\left\{n_m\right\}}\int d\Psi^\dagger d\Psi \,  e^{- \Psi^\dagger \Psi} \,\, ^L\hspace{-0.1cm}\braket{n_m|\Psi}^{R\, L}\braket{\Psi|\hat{O}|n_m}^R\\
		&=\int d\Psi^\dagger d\Psi \,  e^{- \Psi^\dagger \Psi} \,\, ^L\hspace{-0.1cm}\bra{\pm \Psi}\hat{O} \underbrace{\sum\limits_{\left\{n_m\right\}}\ket{n_m}^{R\, L}\bra{n_m}}_{\hat{1}}\ket{\Psi}^R\\
		&=\int d\Psi^\dagger d\Psi \,  e^{- \Psi^\dagger \Psi} \,\, ^L\hspace{-0.1cm}\braket{\pm \Psi|\hat{O}|\Psi}^R,
	\end{split}
\end{align}
where $\Psi$ is composed of Grassmann variables for fermions and is a vector of complex number for bosons \cite{stoof2009ultracold}. 

\subsection{Exact expresssions for $G(r, \tau)$ for a free system}\label{subsec_field_theory}

The Fourier transform of the Green's function
\begin{equation}
	\widetilde{G}_n=\frac{1}{-i \hbar\omega_n-\mu+\mathbb{H}}=\sum\limits_{m} \phi_m^{R} {\phi_m^{L}}^\dagger\frac{1}{-i \hbar\omega_n+\zeta_m},
	\label{eq_G_n}
\end{equation}
determines $G(r, \tau)$.

One can perform the Matsubara sums exactly (done in Mathematica \cite{Mathematica}) in $\widetilde{G}_n$ to obtain
\begin{eqnarray}
&&G\left(\tau\right)=\sum\limits_{n=-\infty}^{\infty} \widetilde{G}_n e^{-i \omega_n \tau}=\sum\limits_{m}\phi_m^{R} {\phi_m^{L}}^\dagger\sum\limits_{n=-\infty}^{\infty}\frac{1}{-i \hbar\omega_n+\zeta_m} e^{-i \omega_n \tau}\nonumber\\
&&=\begin{cases}
		-\frac{i}{2\pi}\beta\sum\limits_{m}\phi_m^{R} {\phi_m^{L}}^\dagger \left[\chi^{-2}\Phi_{HL}\left(\chi^{-2}, 1, 1-iz_m\right)-\Phi_{HL}\left(\chi^{2}, 1, i z_m\right)\right]\equiv G^{B}(\tau)
		\vspace{0.5cm}\\
		\frac{\beta}{2\pi }\sum\limits_{m}\phi_m^{R} {\phi_m^{L}}^\dagger\frac{\chi^{-1} {_2F_1}\left(1, \frac{1}{2}-i z_m,\frac{3}{2}-iz_m,\chi^{-2}\right)}{i/2 +z_m}+\frac{\chi {_2F_1}\left(1, \frac{1}{2}+i z_m,\frac{3}{2}+i z_m,\chi^{2}\right)}{-i/2+z_m}\equiv G^{F}(\tau)
		\label{eq_G_tau}
	\end{cases}
\end{eqnarray}
where $\chi=\exp(i \pi \tau/\beta)$, $z_m=\beta \zeta_m/(2pi) $, $\Phi_{HL}$ is the Hurwitz-Lerch zeta function, $_2F_1$ is the hypergeometric function and we introduced the notation $G^{B}$ and $G^{F}$ to denote bosonic ($n_M$ even) and fermionic ($n_M$ odd) functions. Notice that in the presence of poles these functions also reduce to the form
\begin{eqnarray}
	G\left(\tau\right)&\approx &\beta\sum\limits_{m} \phi_m^{R} {\phi_m^{L}}^\dagger \left(\text{Re}{\zeta_m}\right)^{-1} \, e^{-i\omega_{n}\tau},
	\label{eq_G_ress}
\end{eqnarray}
as in the main text. 

\subsection{iTC for bosons}\label{app_G}

\begin{figure*}[!ht]
	\centering
	\includegraphics[width=\linewidth]{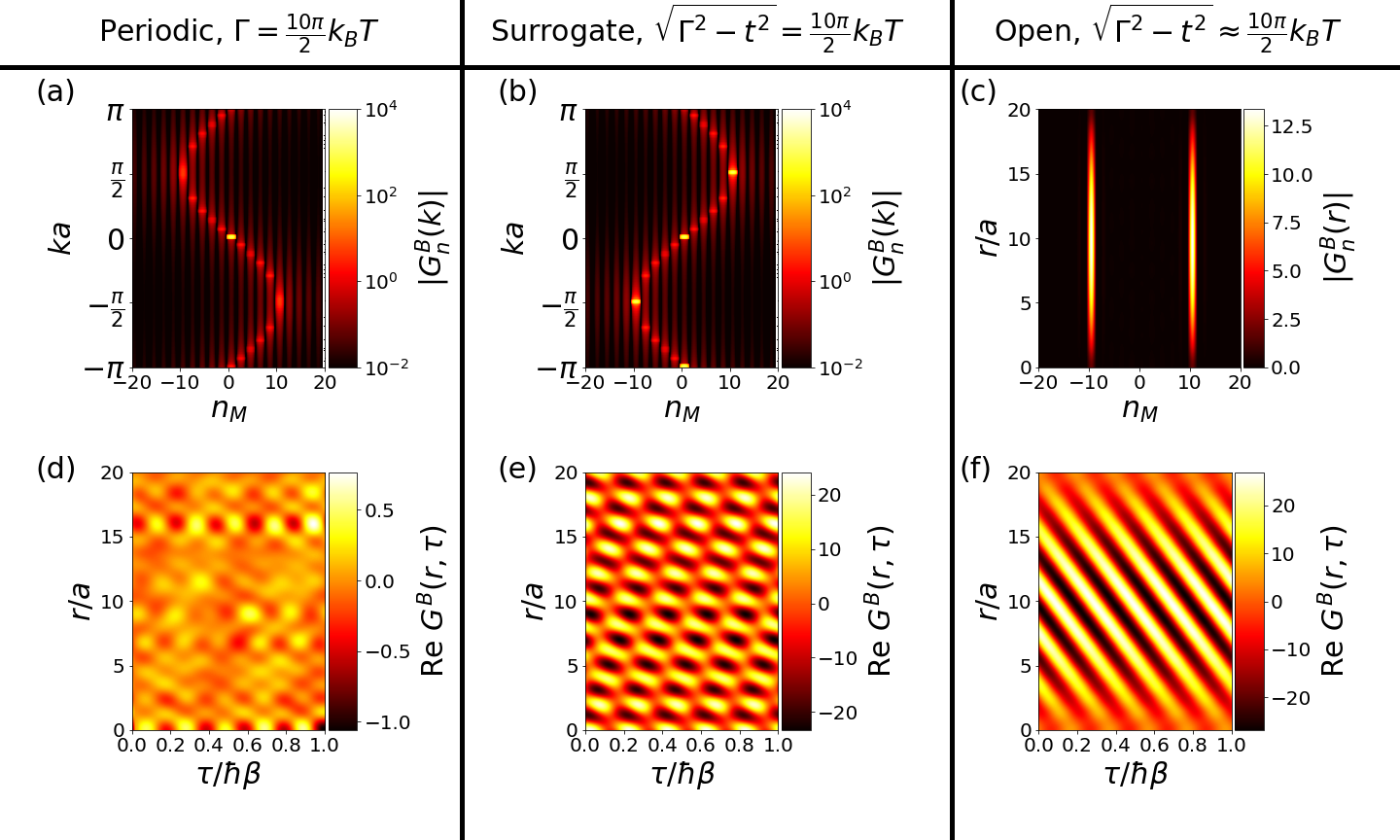}
	\caption{Green's function and its FT for the bosonic Hatano-Nelson model at resonance with $n_M=\pm 10$. Results are for PBC, for the surrogate Hamiltonian, and for OBC, for $t=0.1$, $\mu=-2t-10^{-3}$ (PBC) or $\mu=-10^{-3}$ (OBC and surrogate), and $\beta=1$. We choose $\Gamma$ such that the resonance condition is met (see main text) for all these boundary conditions. A different choice of $\beta$ will only lead to a rescaling of $G$.We start by showing $|\widetilde{G}_n|$ as a function of the Matsubara mode $n_M$ and $k$ (distance to the left edge) for the periodic (open) system. A logscale is used for the periodic systems, such that the features besides the peaks of these functions are visible. This function is shown in (a) for the PBC system, in (b) for the surrogate Hamiltonian, and in (c) for the OBC system. Next, we show the real part of $G(r, \tau)$ as a function of imaginary time $\tau$ and distance $r$, where the spatial dependence is either computed using a FT in momentum, in the case of the periodic systems, or is computed from the wavefunctions, for the open system. The results are shown in (d) for PBC, in (e) for the surrogate Hamiltonian, and in (f) for the OBC system. The results for the imaginary part are similar. We use $100$ $k$-points for the periodic system and a lattice with $20$ sites for the open system. }
	\label{fig_G_Boson_10pi}
\end{figure*}

\begin{figure*}[!htb]
	\centering
	\includegraphics[width=\linewidth]{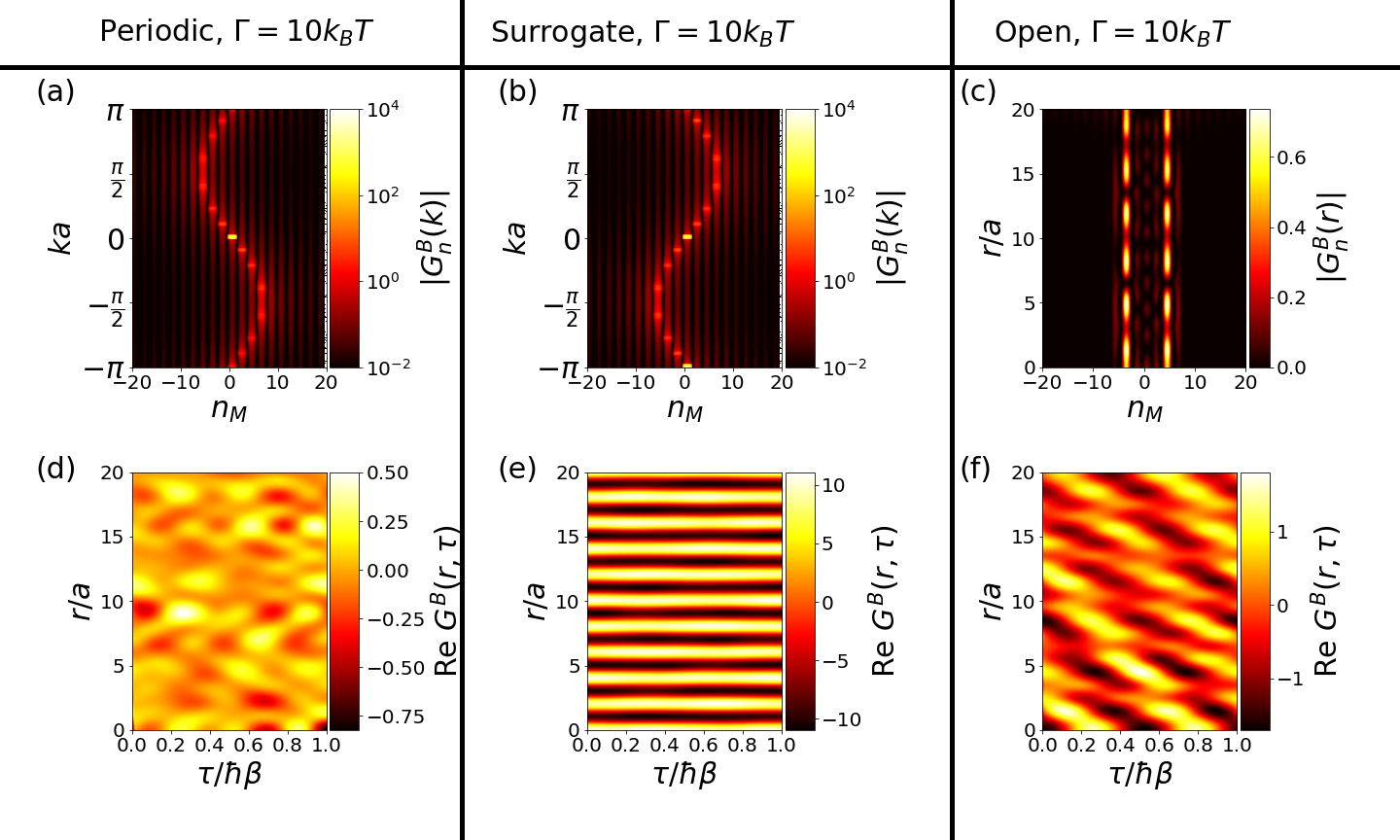}
	\caption{Green's function and its FT for the bosonic Hatano-Nelson model for a large, but non-resonant $\Gamma$. Results are for PBC, for the surrogate Hamiltonian, and for OBC, for $t=0.1$, $\mu=-2t-10^{-3}$ (PBC) or $\mu=-10^{-3}$ (OBC and surrogate), $\Gamma=10 k_BT$ ($\beta \Gamma=10$) and $\beta=1$. A different choice of $\beta$ will only lead to a rescaling of $G$.We start by showing $|\widetilde{G}_n|$ as a function of the Matsubara mode $n_M$ and $k$ (distance to the left edge) for the periodic (open) system. A logscale is used for the periodic systems, such that the features besides the peaks of these functions are visible. This function is shown in (a) for the PBC system, in (b) for the surrogate Hamiltonian, and in (c) for the OBC system. Next, we show the real part of $G(r, \tau)$ as a function of imaginary time $\tau$ and distance $r$, where the spatial dependence is either computed using a FT in momentum, in the case of the periodic systems, or is computed from the wavefunctions, for the open system. The results are shown in (d) for PBC, in (e) for the surrogate Hamiltonian, and in (f) for the OBC system. The results for the imaginary part are similar. We use $100$ $k$-points for the periodic system and a lattice with $20$ sites for the open system.}
	\label{fig_G_Boson_10}
\end{figure*}

\begin{figure*}[!ht]
	\centering
	\includegraphics[width=\linewidth]{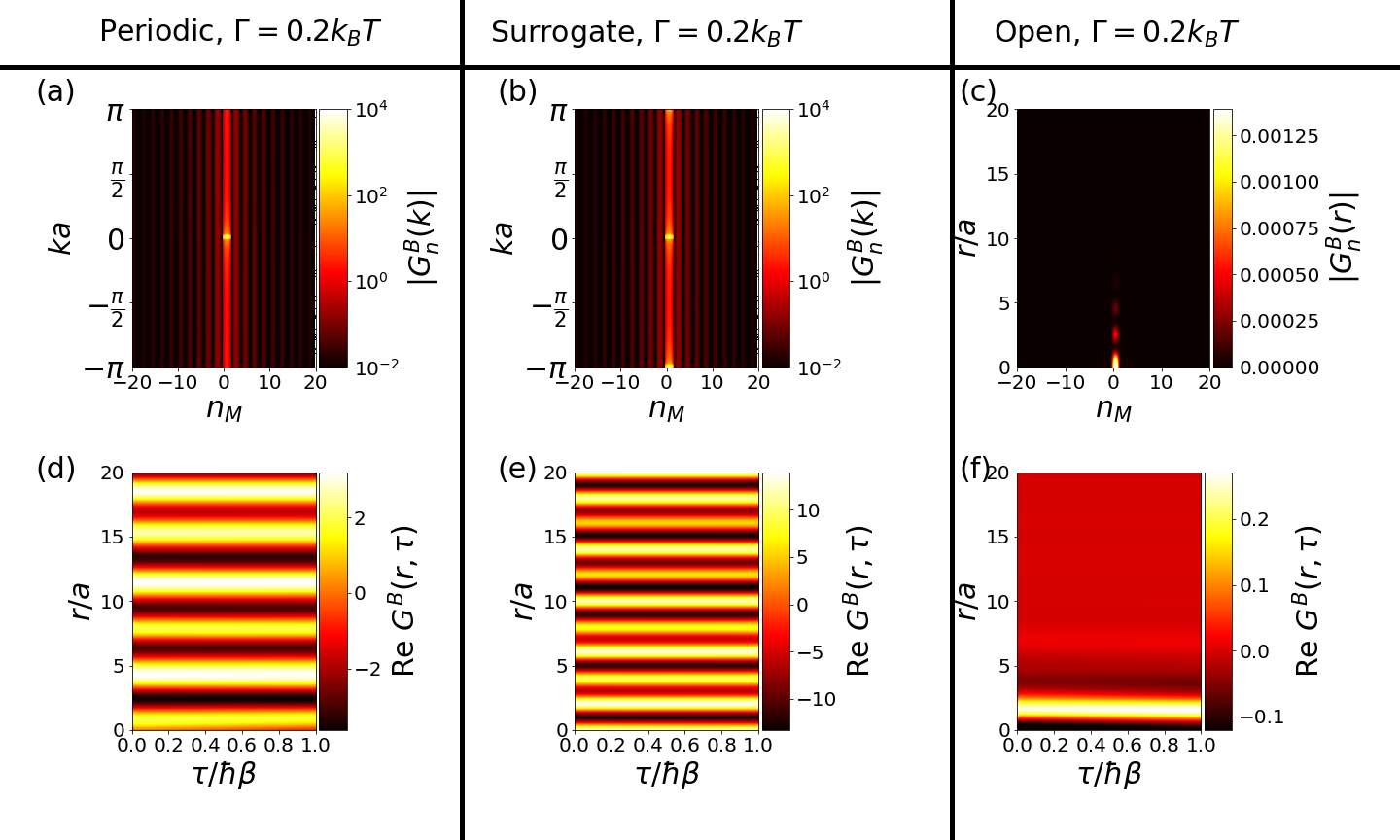}
	\caption{Green's function and its FT for the bosonic Hatano-Nelson model for small $\Gamma$. Results are for PBC, for the surrogate Hamiltonian, and for OBC, for  $t=0.1$, $\mu=-2t-10^{-3}$ (PBC) or $\mu=-10^{-3}$ (OBC and surrogate), $\Gamma=0.2 k_BT$ ($\beta\Gamma=0.2 $) and $\beta=1$. A different choice of $\beta$ will only lead to a rescaling of $G$.We start by showing $|\widetilde{G}_n|$ as a function of the Matsubara mode $n_M$ and $k$ (distance to the left edge) for the periodic (open) system. A logscale is used for the periodic systems, such that the features besides the peaks of these functions are visible. This function is shown in (a) for the PBC system, in (b) for the surrogate Hamiltonian, and in (c) for the OBC system. Next, we show the real part of $G(r, \tau)$ as a function of imaginary time $\tau$ and distance $r$, where the spatial dependence is either computed using a FT in momentum, in the case of the periodic systems, or is computed from the wavefunctions, for the open system. The results are shown in (d) for PBC, in (e) for the surrogate Hamiltonian, and in (f) for the OBC system. The results for the imaginary part are similar. We use $100$ $k$-points for the periodic system and a lattice with $20$ sites for the open system.}
	\label{fig_G_Boson_01}
\end{figure*}

In this Section, we show the Green's function for the bosons. The fact that such systems have peaks for $n_M=0$ make that purely spatial oscillations appear together with the plane waves in imaginary time, seen for the fermionic systems. We analyze $\widetilde{G}_n^B$ and $G(\tau, r)^B$ for $\Gamma$: (i) resonant, Supplementary Figure~\ref{fig_G_Boson_10pi}; (ii) off resonant and larger than $k_BT$, Supplementary Figure~\ref{fig_G_Boson_10}; and off resonant and smaller than $k_BT$, Supplementary Figure~\ref{fig_G_Boson_01}.  

The $\left|\widetilde{G}_n\right|$ for PBC at a resonance (Supplementary Figure~\ref{fig_G_Boson_10pi}), presents a sequence of small peaks together with a very bright peak at $n_M=0$ and $k=0$. For the surrogate, there will be peaks at $n_M=\pm 10$ and $k=\pm \pi/2$, together with the ones at $n_M=0$ for $k=0$ or $k=\pi$. For OBC, there are only peaks at $n_M=\pm 10$, without any signature at $n_M=0$. This is a finite-size effect, as small systems have a gap of order $\mathcal{O}\left(1/M\right)$. As the case for the fermions, the pole structure of $\left|\widetilde{G}_n\right|$ explains the features in $G(r,\tau)$, with the PBC system having an amorphous behavior, the surrogate showing an interference pattern, and the OBC system showing behavior typical of a wavefront. 

For large, non-resonating $\Gamma$ (Supplementary Figure~\ref{fig_G_Boson_10}), the system presents pronounced peaks only at $n_M=0$ for PBC and surrogate, and at $n_m=\pm 4$ for the OBC. The fact that the peak at $n_M=0$ occurs for only $k=0$ for PBC makes that $G(r, \tau)$ has features of an amorphous phase. Conversely, for the surrogate, there is a peak also at $n_M=0$ and $k=\pi/a$, so there will be oscillations on space with $\lambda=2\pi/k=2a$. The presence of the peaks at $n_M=\pm 4$, favor oscillations with $\mathcal{T}=0.5 \hbar \beta$, although they are not ordered because there is a significant weight on other modes.

For small, non-resonating $\Gamma$ (Supplementary Figure~\ref{fig_G_Boson_01}), there are pronounced peaks only for $n_M=0$. For the PBC and the surrogate, the weight of the poles is distributed among different values of $k$. For OBC, there are oscillations in space but a very quick exponential decay of the correlation function. There is modulation in space for PBC and surrogate, although no predominant wavelength is observed. For OBC, there is only decay in space.

\clearpage
\renewcommand{\refname}{Supplementary References}
\printbibliography[heading=subbibliography]


\end{document}